\documentclass[times]{stvrauth}
\usepackage[utf8]{inputenc}
 \usepackage{graphicx}
\usepackage{setspace}
\usepackage{adjustbox}
\usepackage{cite}
\usepackage{amsmath,amssymb,amsfonts, pifont}
\usepackage{algorithmic}
\usepackage{graphicx}
\usepackage{textcomp}
\usepackage{xcolor}
\usepackage{listings}
\usepackage{color}
\usepackage{anyfontsize}
\usepackage{multirow}
\usepackage{lastpage}

\usepackage{array}
\usepackage[section]{placeins}

\usepackage{graphicx}
\usepackage{listings}
\definecolor{lightgray}{rgb}{9, 9, .9}
\definecolor{darkgray}{rgb}{.4, .4, .4}
\definecolor{purple}{rgb}{0.65, 0.12, 0.82}
\definecolor{gray}{rgb}{0.4,0.4,0.4}
\definecolor{line-numbers}{rgb}{0.4,0.4,0.4}
\definecolor{tags}{rgb}{1, 0, 0}
\definecolor{darkblue}{rgb}{0.0,0.0,0.6}
\definecolor{cyan}{rgb}{0.0,0.6,0.6}
\definecolor{highlight}{HTML}{ffffff}

\usepackage{soul}

\newcommand{\hlc}[2][yellow]{{%
    \colorlet{foo}{#1}%
    \sethlcolor{foo}\hl{#2}}%
}

\lstdefinelanguage{Groovy}
{
  sensitive=true,%
    morecomment=[l]//,%
  morecomment=[s]{/}{/},%
  morestring=[b]",%
  morestring=[b]',%
  stringstyle=\color{black},
  identifierstyle=\color{darkblue},
  keywordstyle=\color{cyan},
  morekeywords={abstract,any,as,boolean,break,byte,case,catch,char,
  class, const,continue,def,default,do,double,else,extends,false,final,finally, float,for,goto,if,implements,import,instanceof,in,int,interface,label, long,native,new,null,package,private,protected,public,return,short, static,strictfp,super,switch,synchronized,this,throw,throws,transient, true,try,void,volatile,while,with}
}
\usepackage[T1]{fontenc}
\usepackage{listings,xcolor,beramono}
\usepackage{tikz}
\makeatletter
\newenvironment{btHighlight}[1][]
{\begingroup\tikzset{bt@Highlight@par/.style={#1}}\begin{lrbox}{\@tempboxa}}
{\end{lrbox}\bt@HL@box[bt@Highlight@par]{\@tempboxa}\endgroup}

\newcommand\btHL[1][]{%
  \begin{btHighlight}[#1]\bgroup\aftergroup\bt@HL@endenv%
}
\def\bt@HL@endenv{%
  \end{btHighlight}%
  \egroup
}
\newcommand{\bt@HL@box}[2][]{%
  \tikz[#1]{%
    \pgfpathrectangle{\pgfpoint{1pt}{0pt}}{\pgfpoint{\wd #2}{\ht #2}}%
    \pgfusepath{use as bounding box}%
    \node[anchor=base west, fill=orange!30,outer sep=0pt,inner xsep=1pt, inner ysep=0pt, rounded corners=3pt, minimum height=\ht\strutbox+1pt,#1]{\raisebox{1pt}{\strut}\strut\usebox{#2}};
  }%
}
\makeatother

\lstdefinestyle{Groovy}{
    language={Groovy}, 
     moredelim=**[is][\btHL]{`}{`},
    moredelim=**[is][{\btHL[fill=green!30]}]{*}{*},
    moredelim=**[is][{\btHL[fill=cyan!30]}]{~}{~},
    extendedchars=true,
        }
\usepackage{qtree}
\usepackage{multicol}        
\usepackage{moreverb}

\usepackage[colorlinks,bookmarksopen,bookmarksnumbered,citecolor=red,urlcolor=red]{hyperref}

\newcommand\BibTeX{{\rmfamily B\kern-.05em \textsc{i\kern-.025em b}\kern-.08em
T\kern-.1667em\lower.7ex\hbox{E}\kern-.125emX}}

\def\volumeyear{2010}

\begin{document}

\runningheads{~M.~Alalfi.~et al.}{A Mutation Framework for Evaluating Security Analysis tools in IoT Applications}

\title{A Mutation Framework for Evaluating Security Analysis tools in IoT Applications}

\author{Manar H. Alalfi\corrauth, Sajeda Parveen, Bara' Nazzal}

\address{Ryerson university, Department of Computer Science, Toronto, Canada}

\corraddr{Ryerson university, Department of Computer Science, Toronto, Canada}

\begin{abstract}
	With the growing and widespread use of Internet of Things (IoT) in our daily life, its security is becoming more crucial. To ensure information security, we require better security analysis tools for IoT applications. Hence, this paper presents an automated framework to evaluate taint-flow analysis tools in the domain of IoT applications. First, we propose a set of mutational operators tailored to evaluate three types of sensitivity analysis, flow, path and context sensitivity. Then we developed mutators to automatically generate mutants for those types. We demonstrated the framework on a subset of mutational operators to evaluate three taint-flow analyzers, SaINT, Taint-Things and FlowsMiner. Our framework and experiments ranked the taint analysis tools according to precision and recall as follows: Taint-Things (99\% Recall, 100\% Precision), FlowsMiner (100\% Recall, 87.6\% Precision), and SaINT (100\% Recall, 56.8\% Precision). To the best of our knowledge, our framework is the first framework to address the need for evaluating taint-flow analysis tools and specifically those developed for IoT SmartThings applications.
\end{abstract}


\maketitle

	\footnotetext[2]{
\href{https://github.com/SajedaParveen/A-Mutation-Framework-for-Evaluating-Security-Analysis-tools-in-IoT-Applications}{\texttt{Dataset produced by the proposed mutation framework}}

\href{https://cresset.scs.ryerson.ca/Mutations}{\texttt{Mutation Framework Webfront}}}



\section{Introduction}\label{Introduction}
\vspace{-2pt}

Internet of Things (IoT) is a new revolutionary technological addition of this era which has the capability to connect everything and everyone.
IoT is a collection of multiple interconnected objects, services, humans, and devices that can communicate and share data and information to achieve a common goal in different areas and applications \cite{mahmoud2015internet}. IoT is impacting our day-to-day life. It can be used in different industries like transportation, infrastructure applications, agriculture, healthcare, energy production, manufacturing, distribution, or even in areas such as smart homes or elder care. IoT covers an expansive area of application and is gradually becoming an integral part of our lives. Thus, its security is a crucial challenge to deal with. 

Security analysis is the process of detecting security vulnerabilities to mitigate or remove them in order to make a system or an application more secure. Security analysis for IoT applications is a relatively new and growing area of research. New tools were developed to identify security vulnerabilities in this domain, however, there is a lack of evaluation frameworks to quantitatively test the quality of these tools. Evaluation is the process that allows us to make sure our solution does the job it has been designed for and to think about how it could be improved \cite{Evaluatingsolutions}.  When this evaluation process is done by a framework it is called evaluation framework. Mutation testing was successfully used for evaluation in other domains such as evaluating clone detection tools \cite{roy2009mutation}, software testing systems for Fortran \cite{king1991fortran}, gene regulation models \cite{watson2004towards}, software system clustering \cite{doval1999automatic}, and inter-class mutation for Java \cite{ma2002inter}.
  
Mutation testing is a fault-based software testing technique \cite{jia2010analysis}, which mimics errors that are often made by programmers. 
Artificial errors can be injected into the main application, where each generated file, known as a mutant, only includes one change. Mutants are run against test cases to check on how many mutants were identified or, in mutation testing terminology, killed. The ratio of killed mutants over the total number of mutants is then computed, which helps to identify better test cases and less error-prone systems. Although these mutants are generated and executed efficiently by automated methods, around 20-30\% of the mutants are functionally equivalent to the original program and are not useful for testing \cite{offutt1994using}. When the ratio is computed, it is computed based on the nonequivalent mutants. 
 
SmartThings \cite{SamsungSmartThings} is one of the most popular SmartHome platforms for IoT, other platforms include, Apple HomeKit \cite{AppleHomeKit}, Amazon Web Services (AWS) \cite{AmazonWebServices}, IBM Watson \cite{IBMWatson}, etc. In this paper, we only focus on Samsung's SmartThings applications since it is a popular platform with a large developer community that provides support and has numerous open-source official and third-party applications available that can be used for analysis. In addition, most of the recent security analysis tools that target IoT applications were developed for SmartThings applications. These security analysis tools use static taint flow analysis. Static taint flow analysis is a technique where we try to find a path between user controllable input to vulnerable function without running the program. A study conducted by Celik et al. \cite{celik2018sensitive} concluded that IoT platforms use similar programming structures and the differences lie only in the communication protocols between IoT devices and edge systems. As such, we believe that the approach we discuss in this paper can be generalized to other IoT platforms.

Since the field of IoT is still growing, the available testing tools might be limited. For example some IoT platforms are new and the availability of datasets for testing might be restricted or they might not be large enough. Furthermore, specific problems might not have enough test cases to adequately make conclusions. This is one of the areas a mutational framework can be helpful by automatically generating mutation cases that can be tested instead. 

Another problem is that different tools might have different approaches for handling the available tests, making it difficult to directly compare them in a quantitative way, since they may not have a shared basis for the comparison. A mutational framework can help with this, by automating the process and providing uniform tests, it allows us to assess the tools quantitatively in a more meaningful manner.

The primary objective of this study is to build an automated framework that is able to quantitatively evaluate static security analysis tools in the field of IoT. The primary objective of this paper is to answer the following research questions.
  
\textit{RQ1.} How can we quantitatively evaluate static taint analysis tools in IoT to ensure security?

\textit{RQ2.} \hlc[highlight]{Is the proposed methodology able to differentiate the security analysis tools and identify their weakness?}

In order to answer \textit{RQ1},  we developed a set of mutational operators which create security vulnerability related faults in a program. Mutational operators are nothing but a mechanism of creating known faults. Based on whether the security analysis tool can identify the generated mutants or not we can quantitatively evaluate the tools. 

For answering \textit{RQ2}, we ran all of the mutational operators on a benign data-set and created faulty programs. Then, we used the security analysis tools that were available to examine the faulty programs. After that, we generated the tool-specific results and we did a statistical analysis to find out the tools' quantitative results in different areas. 
 
This paper is an extension of a previously published short paper \cite{parveen2020mutation} where we presented an initial version of our mutation framework, and demonstrated it on one category of mutational operators, flow-sensitive analysis mutational operators. The previously published short paper provided a basic description for the approach and handled the issue of flow sensitivity using 5 mutational operators. The extension expanded the scope to include path and context sensitivity tests and provides an additional 7 mutational operators. We have also expanded the tests to include a newly developed tool, FlowsMiner.

In this extended version we make the following contributions:\begin{itemize}
	\item First, We expanded the discussion about the mutation analysis framework \cite{parveen2020mutation}, which is to the best of our knowledge, the first automated framework that is aimed at quantitatively evaluating tools focused on software security vulnerabilities identification in the domain of IoT applications.
	\item Second, we propose an expanded set of unique mutational operators for one class of security vulnerabilities, tainted-flow vulnerability. This includes, in addition to flow-sensitive discussed previously \cite{parveen2020mutation}, path sensitive and context sensitive analysis mutational operators.
	\item Third, we provide a comprehensive evaluation that demonstrates the framework on the evaluation of three static analysis tools, SaINT \cite{celik2018sensitive}, Tainted-Things \cite{schemild}, and FlowsMiner \cite{DBLP:conf/wcre/NaeemA20}.
\end{itemize}

\section{Background}\label{Background}
\vspace{-2pt}
In this paper, we propose a mutational analysis framework that is aimed at evaluating static taint flow analysis tools for SmartThings. This section includes all the necessary and essential background information that are needed to explain our approach and contribution. 

To understand the structure of SmartThings applications, we provide Listing\ref{list:smartthingsApp} which demonstrates the code structure with an example. It is written in Groovy, a language similar to Java, and has three main sections, the definition, preference and methods sections.
 \begin{lstlisting}[style=Groovy, caption={SmartThings application structure}, label=list:smartthingsApp]
definition(
  name: "SmartApp",
   //.....
)

preferences {
  section("Turn on when motion detected:"){ 
    input "themotion", "capability.motionSensor", 
        required: true, title: "Where?"
  }
}


def installed() {
  initialize()
}

def updated() {
  unsubscribe() 
  initialize()
}

def initialize() {
}
\end{lstlisting}

First, the definition section includes a set of meta-data that is usually populated from the values the developer enters when creating the SmartApp \cite{SmartThingsFirstAPP}. Second, the preferences section includes values provided by users as inputs. When installing the SmartApp, a user will be given a screen to customize these inputs. The input section also includes the access control of a device, which is expressed using a capability model; SmartThings categorizes devices into their capabilities, that is, what the device is capable of \cite{SmartThingsFirstAPP}. Third, the methods section includes predefined methods that are called during the app installation, update, or deletion, while others are event handlers specified in the event subscriptions and any other methods necessary for implementing the SmartApp.

It is necessary to prevent leakage of confidential information as well as keeping the trusted data untainted. There are a lot of studies in this field to quantify the tolerable level for leakage. However, any leakage will make a system vulnerable, the caveat being that there is no perfectly secure system either. So, in our day-to-day life, we often have to face vulnerabilities related to information leakage.

Program analysis refers to the process of analyzing a computer program to look for certain properties, such as its correctness, stability, or security \cite{nielson2015principles}. 
In this paper, we will evaluate some tools used for program analysis to ensure security.

Static analysis is a kind of program analysis techniques that has the ability to reveal the vulnerabilities during the development phase of the program \cite{jovanovic2006pixy}. It analyzes the code itself before the running phase. Static analysis is the core of all the tools used in this study. Dynamic analysis, on the other hand, is a program analysis technique that is done by using data gathered during the run-time. It is capable of presenting an exact picture of a software system as it reveals the system's true nature \cite{4815280}. 
    
Taint Analysis attempts to identify variables that have been \textit{tainted} with user-controllable input and traces them to possible vulnerable functions, also known as  
\textit{sinks} \cite{owaspwebsite}. If the tainted variable gets passed to a sink without first being sanitized, it is flagged as a vulnerability \cite{owaspwebsite}. Input sanitization describes cleansing and scrubbing user input to prevent it from jumping the fence and exploiting security holes \cite{websanitization}. 
Taint Analysis can be classified into the following types, which specify the amount of precision the analysis can provide: 
a) Flow sensitivity takes into account the order of execution when doing program analysis \cite{nielson2015principles}, 
b) Path sensitivity requires that the predicates at conditional branches are considered in a program analysis \cite{nielson2015principles}, and 
c) Context-sensitive analysis spans multiple procedures, considering a target function block within the context of the code calling it \cite{sharir1978two}. We now discuss each of these types in detail.
\subsection{Flow Sensitivity}\label{subsec:Flow Sensitivity}
In program analysis, flow sensitivity assess the order of execution \cite{nielson2015principles}. Listing \ref{list:flowExample} provides an example IoT app code snippet to explain flow sensitivity. A flow-sensitive analysis would say that this app is benign, as the value of the input \textit{\$people} will not be leaked through \textit{messages} variable to the sink method \textit{send}. But a flow-insensitive analysis will flag it as probable leakage, since it can not distinguish the value change for the same variable, \textit{messages}.

\begin{lstlisting}[style=Groovy, caption={Flow Sensitivity Example }, label=list:flowExample]
preferences {
  section("When all of these people leave home") {
    input `"people"`, "capability.presenceSensor", multiple:true
  }
}

def eventHandler(evt) { 
  def messages = ~"people in house "+ `"{$people}"`s~
  messages = *"new event happened"*
  ~send("attackerphone",`messages`)~
}
\end{lstlisting}

\vspace{-0.1 cm}
\subsection{Path Sensitivity}\label{subsec:Path Sensitivity}
Path sensitivity requires that the predicates at conditional branches are considered in a program analysis \cite{nielson2015principles}, an example is illustrated in Listing \ref{list:pathExample}. A path-sensitive analysis will give two different independent paths based on the value of the condition. If the value of the input variable \textit{\$people} is equivalent to zero, it will leak information; however, if the value is not equal to zero, it will have a benign value, and consequently will not leak any sensitive information. A path sensitive analysis will be able to identify precisely which path in the app leaks sensitive data and which does not, whereas path-insensitive analysis will flag the app in Listing \ref{list:pathExample} as tainted, because the path-insensitive analysis cannot identify each path independently. 
   
\begin{lstlisting}[style=Groovy, float, caption={Path Sensitivity Example}, label=list:pathExample]
preferences {
  section("When all of these people leave home") {
    input `"people"`,"capability.presenceSensor",multiple:true
  }
}

def presence(evt) {   
  def messages = "Random"
  def attackerphone = "11111111111"
  if ($people == 0) { 
    messages = `"{$people}"`
  } else {
    messages = *"Random"*
  }
  ~sendSMS(attackerphone, `messages`)~
}

\end{lstlisting}
\subsection{Context Sensitivity}\label{subsec:Context Sensitivity}
Context-sensitive analysis spans multiple procedures and considers a target function block within the context of the code calling it \cite{sharir1978two}. For example, in Listing \ref{list:contextExample} the function \textit{createmsg} is called from function presence with two different kinds of parameters. The first call will be returning a benign value and the 2nd call will return a malicious value. A context-sensitive analyzer will be able to distinguish between this call. So if the \textit{message1} is passed to a sink it will be benign context and if \textit{message2} is passed to sink it will be a malicious one.
However, a context-insensitive analyzer will fail to distinguish between these calls and ended up calling both of the cases malicious. In this example Listing \ref{list:contextExample} will be reported benign by a context-sensitive analyzer, while a context-insensitive analyzer will say it is potentially malicious.

\begin{lstlisting}[style=Groovy, caption={Context Sensitivity Example}, label=list:contextExample]
preferences {
  section("When all of these people leave home") {
    input `"people"`, "capability.presenceSensor", multiple:true
  }
  section("temperature") {
    input "thermostat", "capability.thermostat", title: "Thermostat", multiple: false, required: true
  }
}

def presence(evt) {
  def phone = "11111111111"
  def message1 = *createmsg("random")*
  def message2 = `createmsg("{$people}","random")`
  ~sendSms(phone, `message1`)~
}

def createmsg(msg) {
  ~return "hello"~
}

def createmsg(`msg`, flag) {
  `return msg`
}
\end{lstlisting}

\subsection{Mutation Testing } 
Mutation testing is a fault-based testing technique \cite{jia2010analysis} which is widely used. If we generate the mutation carefully in an appropriate location, it can simulate many test criteria that are hard to  reach. 
It will inject one unique fault per file. Mutants are created to check the designed test cases' efficiency. Test cases are performed over the generated mutants to identify or kill them.
The ratio of killed mutants over the total number of mutants is then computed, which indicates the efficiency of the test cases. Thus, it helps to both identify the good test cases and the required test scenarios. Although these mutants are generated and executed efficiently by automated methods, many of the mutants are functionally equivalent to the original program and are not useful for testing \cite{offutt1994using}. When the ratio is computed to check efficiency, it is computed based on mutants that are not equivalent to the original program. Equivalent mutants are still a hard problem to tackle. Introducing equivalent mutants does not change or modify the actual state of the original program. 

In our work, we  utilize equivalent mutants differently and that to evaluate the tools for their false positive rate. We do so by creating benign mutants which does not change the actual functionality of the program, but the purpose would be to inject benign program constructs that are related to flow, path and context sensitivity. If the tool we evaluate reports those changes as vulnerability then that would impact their score for reporting false positive score.
\subsection{Source transformation technology via TXL}
TXL is a unique programming language specifically designed to support computer software analysis and source transformation tasks \cite{TXL}. To use it, one needs to define a grammar or reuse a predefined one. The grammar helps to parse the input files and based on it, transformation rules can be designed. Grammar and transformation  rules are the core elements of this language. TXL has a pure functional superstructure that provides scoping, abstraction, parametrization, and recursion, over Prolog-like structural rewriting rules which provide pattern search, unification, and implicit iteration \cite{TXL}.

 \section{Survey of Related Work} 
Our mutation-based evaluation framework uses source transformation technology \cite{TXL} to help automate the stage of mutation generation. 

Roy et al. \cite{roy2009mutation} use the same technology to develop a mutational analysis framework to evaluate clone detection tools. \hlc[highlight]{They introduced operators specific to detection of clone types, which include changes with whitespaces, comments and formatting; renaming and replacement of identifiers; modification of line code such as insertion and deletion; and finally reordering and replacement within the program.} Dan et al. \cite{dan2012smt} use TXL to implement a Semantic Mutation Testing (SMT) tool to mutate the semantics of the language so they can capture different category of faults other than those usually discovered by mutating the program source code \cite{clark2013semantic}.  While both papers use TXL, they were used for other applications, purposes and platforms. Dan et al. \cite{dan2012smt} adopted the mutation analysis for platforms using C and \hlc[highlight]{the operators they used are specific to C semantics and controlling floating points.}. Roy et al. \cite{roy2009mutation} designed their framework for evaluating  \hlc[highlight]{clone detection in} platforms using C, Java, and C\# while our paper applies the mutation analysis to \hlc[highlight]{taint detection in the} SmartThings platform. 

Papadakis et al. \cite{DBLP:journals/ac/PapadakisK00TH19} conducted a survey to highlight the benefit of mutation in empirical studies of software testing. The survey sheds light on the challenges of mutation testing and advises on best practices. 

The next six papers are similar to our work in using mutation for security purposes. This application of mutation analysis was also mentioned in Papadakis et al. \cite{DBLP:journals/ac/PapadakisK00TH19}. Avancini and Ceccato \cite{avancini2010towards} presented a preliminary investigation on combining static analysis with a genetic algorithm. Their approach uses mutation to support security testing \hlc[highlight]{by selecting the inputs in the tested page, pairing them, crossing and modifying the parameters. The goal is to test whether the inputs can go through the targeted paths}. Unlike our work, the prototype is designed for web applications using Java \hlc[highlight]{and the focus is on cross site scripting vulnerability} 

Matthis et al. \cite{mathis2017detecting} developed a lightweight mutation-based analysis tool that systematically mutates dynamic values returned by sensitive sources to assess whether the mutation changes the values passed to sensitive sinks. By mutating the values, the authors found flows between sources and sinks \cite{mathis2017detecting}. Their paper uses a dynamic experimental approach alongside mutation, which is different from our method. The proposed prototype by the authors named MUTAFLOW is implemented using Jimple code and Soot \cite{lam2011soot}. The main goal of Matthis et al. \cite{mathis2017detecting} work is to analyze the information flow to assess the security of applications, but our goal, it is to evaluate taint flow analysis tools. 

Loise et al. \cite{loise2017towards} introduced 15 mutation operators for Java to handle security issues. Specifically, their main goal is to design and support security-aware mutational operators for mutation analysis. \hlc[highlight]{The operators they proposed are specific to Java-based web applications and includes mutations to test for handling of cookies, XML parsing, SQL injection, encryption and sanitation.} These operators are not used for evaluating static \hlc[highlight]{taint} analysis tools like our approach. Mouelhi et al. \cite{mouelhiv2008generic} proposed a meta-model that provides a generic representation of access control models security policies and then defined a set of mutation operators at this generic level. In another paper, Mouelhi et al. \cite{mouelhi2009transforming} introduced an approach to reuse existing functional test cases for security testing. Instead of using mutational operators for evaluating tools, these two papers present new methods for mutation analysis of security policy test cases. Dadeau et al. \cite{dadeau2011mutation} \hlc[highlight]{used a mutational approach to test security protocols, rather than app analyzers. They} used mutation of High-Level Security Protocol Language to design a model‐based testing approach. The proposed mutational operators created leaks in security protocols and mimicked implementation errors. The created errors are used to validate the implementation of a security protocol based on an HLPSL model. The jMuHLPSL tool, a prototype tool for this paper, is implemented using JAVA.

Jia and Harman \cite{jia2010analysis} conducted a related study to analyse 
how mutation testing evolved over time. The papers mentioned for security policies in this survey mostly create mutation operators that inject common security-related flaws into different types of platforms. Le Traon et al. \cite{le2007testing} proposed eight mutation operators classified into four categories for the Organization-Based Access Control OrBAC policy. Hwang et al. \cite{hwang2008systematic} provided a solution that applies rule-level and clause-level mutational operators to test firewall policies. Martin and Xie \cite{martin2007fault} presented a fault model for access control policies and a framework to explore it. The fault model is created based on mutation analysis to test XACML, an Oasis standard XML syntax for defining security policies.

The next three papers share the same goal of evaluating taint analysis tools in the Android applications. ReproDroid, implemented by Pauck et al. \cite{pauck2018android}, is a framework that produces an accurate comparison of Android taint analysis tools. The framework adopted the Android App Analysis Query Language (AQL) for precisely formulating questions about app properties, such as their flows, as well as providing a query execution system which can interact with diverse tools and a query wizard for determining the ground truth in apps and for executing the specified benchmarks and rating their outcome. Qiu et al. \cite{qiu2018analyzing} also performed a comparison in three taint analysis tools in standard configuration setup. \hlc[highlight]{To do that, they used a set of benchmark apps.} Bonnet et al. \cite{bonett2018discovering} presented the Mutation-based soundness evaluation $(\mu SE)$ framework, which systematically evaluates Android static analysis tools to discover, document, and fix flaws and that by leveraging the well-founded practice of mutation analysis. \hlc[highlight]{Their approach uses operators to generate tests for cases such as SSL vulnerabilities and data leakage, but the generated tests are specific to Android.}

Klinger et al. \cite{klinger2019differentially} represent the first automated technique for deferentially testing the soundness and precision of program analyzers. To evaluate the six well known developed program analyzers the researchers used tens of thousands of automatically generated benchmarks. \hlc[highlight]{Their approach is specific to C analyzers and uses assert statements to target configured properties such as numerical properties.}

Our research targets static analysis tools for SmartThings IoT apps. Static analysis has wide applications beyond IoT, such as analyzing Android apps. One of the shared problems is handling of the analysis sensitivity. FlowDroid \cite{flowdroid} is one of the first static analysis tools for android that provide context, field, object, and flow sensitivity analysis. To achieve that, it uses Soot to provide an Intermediate Representation (IR) from the Java code and constructs a call graph from it. The analysis is then performed on the call graph. IoT apps analysis can use different approaches, including the recovery of an intermediate representation (IR), and then try to provide different sensitivities for the analysis on the recovered IR, like the approach proposed by Celik et al. \cite{celik2018sensitive}. We try to evaluate a set of such tools and their ability to provide flow, path and context sensitive analysis and that using our mutational framework, presented in this paper, \hlc[highlight]{which automatically generates mutants specifically targeting SmartThings platform}. 

In Table \ref{fig:RelatedWorks}, we provide a summary of the surveyed papers, the goal of the design, the targeted platform and the methods they used.

\begin{table}[h]
\centering
\caption{Related work comparison}
\label{fig:RelatedWorks}
\begin{adjustbox}{width=1.1\textwidth,center}
\begin{tabular}{|l|l|l|l|l|l|l|l|} 
\hline
\textbf{Paper}                                                                     & \textbf{Platform}                                                                           & \begin{tabular}[c]{@{}l@{}}\textbf{Method}\\\textbf{used in the}\\\textbf{approach}\end{tabular} & \begin{tabular}[c]{@{}l@{}}\textbf{Tool}\\\textbf{ produced}\end{tabular}          & \begin{tabular}[c]{@{}l@{}}\textbf{Goal of the}\\\textbf{ approach}\end{tabular}                                                    & \begin{tabular}[c]{@{}l@{}}\textbf{Language or}\\\textbf{ platform}\end{tabular}      & \begin{tabular}[c]{@{}l@{}}\textbf{Evaluating}\\\textbf{ other tools}\end{tabular}        & \textbf{Designed for}                                                                                                                          \\ 
\hline
\begin{tabular}[c]{@{}l@{}}Avancini\\ and Ceccato\\\cite{avancini2010towards}\end{tabular}                     & \begin{tabular}[c]{@{}l@{}}Web\\ application\end{tabular}                                   & \begin{tabular}[c]{@{}l@{}}Static \\analysis and\\ mutation\end{tabular}                         & Prototype tool                                                                     & Security testing                                                                                                                    & Java                                                                                  &                                                                                           & \begin{tabular}[c]{@{}l@{}}Support security testing\\by generating proper\\ test cases for actual\\ vulnerabilities\end{tabular}               \\ 
\hline
\begin{tabular}[c]{@{}l@{}}Matthis et al.\\\cite{mathis2017detecting} \end{tabular}                                                                        & Android                                                                                     & \begin{tabular}[c]{@{}l@{}}Dynamic\\experimental\\approach\\and mutation\end{tabular}            & \begin{tabular}[c]{@{}l@{}}MUTAFLOW\\ prototype\end{tabular}                       & \begin{tabular}[c]{@{}l@{}}Assessing the\\ security of\\ applications\end{tabular}                                                  & \begin{tabular}[c]{@{}l@{}}Jimple code\\ and Soot\end{tabular}                        &                                                                                           & \begin{tabular}[c]{@{}l@{}}Analyzing information\\ flow by dynamically\\ mutating sensitive\\source values\end{tabular}                        \\ 
\hline
\begin{tabular}[c]{@{}l@{}}Loise et al.\\\cite{loise2017towards}\end{tabular}                                                                       & Java                                                                                        & Mutation                                                                                         & \begin{tabular}[c]{@{}l@{}}Only\\ mutational\\ operators\end{tabular}              & Security testing                                                                                                                    & \begin{tabular}[c]{@{}l@{}}PIT (Parallel \\Isolated Test)\end{tabular}                & \begin{tabular}[c]{@{}l@{}}Can be used to\\ evaluate static\\ analysis tools\end{tabular} & \begin{tabular}[c]{@{}l@{}}Design and support\\ security aware\\ mutational operators\end{tabular}                                             \\ 
\hline
\begin{tabular}[c]{@{}l@{}}Mouelhi et al.\\ mouelhi2009\\ transformin\\ \cite{mouelhi2009transforming} \end{tabular} & Java                                                                                        & \begin{tabular}[c]{@{}l@{}}Mutation\\and dynamic\\ analyses\end{tabular}                         & \begin{tabular}[c]{@{}l@{}}Proposed a\\method and \\tools\end{tabular}             & \begin{tabular}[c]{@{}l@{}}Security Policy\\ testing\end{tabular}                                                                   & \begin{tabular}[c]{@{}l@{}}AOP, model\\ transformation\end{tabular}                   &                                                                                           & \begin{tabular}[c]{@{}l@{}}Mutation analysis\\ of Security Policies\\ test cases\end{tabular}                                                  \\ 
\hline
\begin{tabular}[c]{@{}l@{}}Mouelhi et al.\\ mouelhi2008\\ generic\\\cite{mouelhiv2008generic}\end{tabular}     & \begin{tabular}[c]{@{}l@{}}All rule-based\\ formalism\end{tabular}                          & Mutation                                                                                         & \begin{tabular}[c]{@{}l@{}}Mutation Tool\\ generated\end{tabular}                  & \begin{tabular}[c]{@{}l@{}}Security Policy\\ testing\end{tabular}                                                                   & Kermeta                                                                               &                                                                                           & \begin{tabular}[c]{@{}l@{}}Mutation analysis\\ of Security\\Policies test cases\end{tabular}                                                   \\ 
\hline
\begin{tabular}[c]{@{}l@{}}Dadeau et al.\\\cite{dadeau2011mutation}\end{tabular}                                                                      & \begin{tabular}[c]{@{}l@{}}High-Level\\ Security Protocol\\ Language\\ (HLPSL)\end{tabular} & Mutation                                                                                         & \begin{tabular}[c]{@{}l@{}}jMuHLPSL tool,\\ a prototype tool\end{tabular}          & \begin{tabular}[c]{@{}l@{}}Security Protocols\\ testing\end{tabular}                                                                & Java                                                                                  &                                                                                           & \begin{tabular}[c]{@{}l@{}}A testing technique\\ to validate\\an implementation\\ of security protocol,\\ based on a HLPSL model\end{tabular}  \\ 
\hline
\begin{tabular}[c]{@{}l@{}}Pauck et al.\\\cite{pauck2018android}\end{tabular}                                                                       & Android                                                                                     & Taint analysis                                                                                   & \begin{tabular}[c]{@{}l@{}}ReproDroid,\\ a framework\end{tabular}                  & \begin{tabular}[c]{@{}l@{}}Evaluate taint-flow\\ analysis tools\end{tabular}                                                        & \begin{tabular}[c]{@{}l@{}}Android App\\ Analysis Query\\ Language (AQL)\end{tabular} & \begin{tabular}[c]{@{}l@{}}Evaluate static\\ analysis tools\end{tabular}                  & \begin{tabular}[c]{@{}l@{}}Evaluate static\\ analysis tools\end{tabular}                                                                       \\ 
\hline
\begin{tabular}[c]{@{}l@{}}Qiu et al.\\ \cite{qiu2018analyzing} \end{tabular}                                                                        & Android                                                                                     & \begin{tabular}[c]{@{}l@{}}Static taint\\ analysis\end{tabular}                                  & No/Experiment only                                                                 & \begin{tabular}[c]{@{}l@{}}evaluate taint-flow\\ analysis tools\end{tabular}                                                        & \begin{tabular}[c]{@{}l@{}}N/A. Done in\\Linux\\environment\end{tabular}              & \begin{tabular}[c]{@{}l@{}}Evaluate static\\ analysis tools\end{tabular}                  & \begin{tabular}[c]{@{}l@{}}Evaluate static\\ analysis tools\end{tabular}                                                                       \\ 
\hline
\begin{tabular}[c]{@{}l@{}}Dan et al. \\\cite{dan2012smt}\end{tabular}                                                                        & C                                                                                           & Mutation                                                                                         & \begin{tabular}[c]{@{}l@{}}Semantic \\Mutation\\ Testing (SMT)\\ Tool\end{tabular} & \begin{tabular}[c]{@{}l@{}}Semantic Mutation\\ Testing\end{tabular}                                                                 & TXL                                                                                   &                                                                                           & \begin{tabular}[c]{@{}l@{}}Developing a \\Semantic Mutation\\ Testing tool\end{tabular}                                                        \\ 
\hline
\begin{tabular}[c]{@{}l@{}}Roy et al.\\\cite{roy2009mutation} \end{tabular}                                                                             & \begin{tabular}[c]{@{}l@{}}C, Java \\and C\#\end{tabular}                                   & Mutation                                                                                         & Framework                                                                          & Clone detection                                                                                                                     & TXL                                                                                   & \begin{tabular}[c]{@{}l@{}}Evaluating\\ Code Clone Detection\\ Tools\end{tabular}         & \begin{tabular}[c]{@{}l@{}}Evaluating Code\\ Clone Detection\\ Tools\end{tabular}                                                              \\ 
\hline
\begin{tabular}[c]{@{}l@{}}Bonnet et al.\\\cite{bonett2018discovering}\end{tabular}                                                                       & Android                                                                                     & Mutation                                                                                         & Framework                                                                          & \begin{tabular}[c]{@{}l@{}}Evaluate Android\\ static analysis tools\end{tabular}                                                    & Java                                                                                  & \begin{tabular}[c]{@{}l@{}}Evaluates Android\\ static analysis tools\end{tabular}         & \begin{tabular}[c]{@{}l@{}}Evaluate Android\\ static analysis\\ tools\end{tabular}                                                             \\ 
\hline
\begin{tabular}[c]{@{}l@{}}Klinger et al.\\\cite{klinger2019differentially} \end{tabular}                                                                     &                                                                                             & benchmark                                                                                        &                                                                                    & \begin{tabular}[c]{@{}l@{}}Evaluated the\\ implications of\\ soundness and\\ precision issues\\ of program\\ analyzers\end{tabular} &                                                                                       &                                                                                           & \begin{tabular}[c]{@{}l@{}}Evaluated the\\ implications of\\ soundness and\\ precision issues\\ of program analyzers\end{tabular}              \\
\hline
\end{tabular}
\end{adjustbox}
\end{table}

 
  \section{Framework architecture} \label{sec:Framework architecture}
\begin{figure}[!t]
	\centering
	\includegraphics[width=1\textwidth]{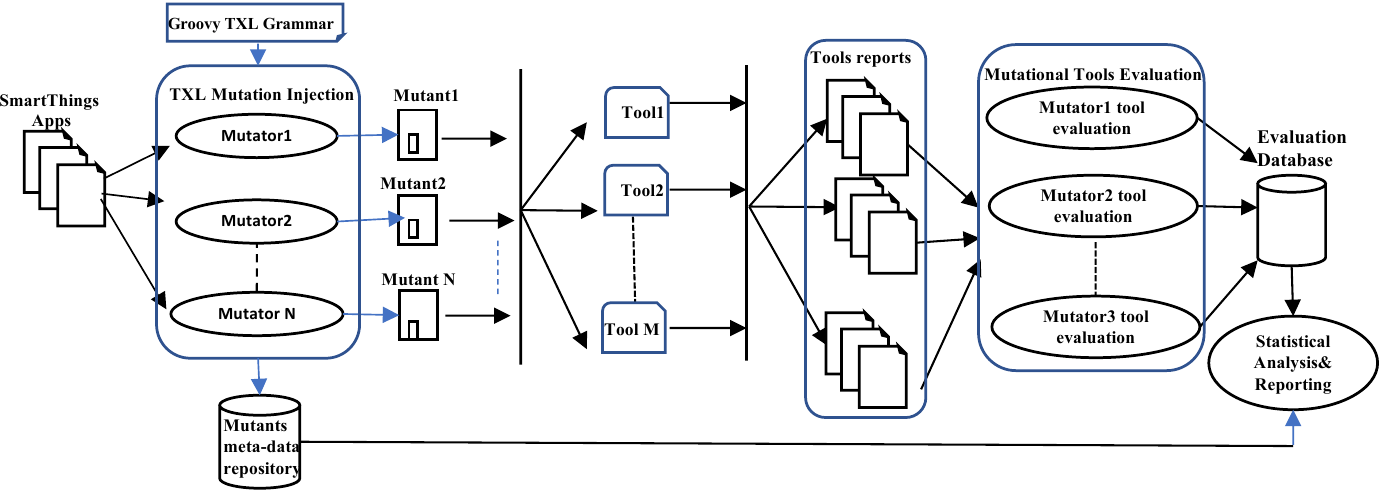}
	\caption{Proposed Mutational Framework}
	\label{fig:ProposedFramework}
\end{figure}

The proposed mutational framework, designed to evaluate tools used for the security analysis of SmartThings apps, is presented in Figure \ref{fig:ProposedFramework}. The framework comprises three main steps which we will explain in the following subsections. Subsection \ref{subsec:autoMutation} introduces the idea of automated mutation generation. Subsection \ref{subsec:EvaluationTools} sheds the light on generation of tool-specific reports. Subsection \ref{subsec:StatisticalAnalysischap4} gives an overview of how the statistical analysis or the quantitative evaluation will be done.
    
\subsection{Automated Mutation Generation}\label{subsec:autoMutation}
Inputs from SmartThings apps are used to generate mutants automatically. The mutational operators will make any existing sink potentially leak these input through different language structure such as:  flow, path, or context (functions). We used the TXL Groovy grammar from Schemild et al. \cite{schemild}  to help automate the mutation generation process. We redefined some parts of the TXL Groovy grammar to accommodate our needs. The TXL Groovy grammar is used to parse and inject one fault/mutation into each file. As an output of this automated injection, all applicable types of mutants for each application are created. An additional output of this step is the storage of all mutant related data into the meta-data repository for future use. All the different kinds of mutational operators used in this step are introduced next.
Our framework proposes 13 new mutational operators. Our aim is to create mutants to evaluate three levels sensitivity analysis: flow, path, and context sensitivity analysis. Flow-sensitivity has six operators whereas path and context each have four and three operators, respectively. We exploited three commonly used sinks in the field of IoT to create the operators, theses sinks are: message, push massage and httpPost. Sensitive information or sensitive input is the user controllable inputs. 

Providing coverage for all the test cases is important, since a tool might be missing a sink or handling it inadequately. By providing a uniform mutational framework that generates all the test cases automatically, it can help test the tools fully even for minor errors. 

We propose the mutational operators to test the main functionalities. Considering that leakage in SmartThings can happen through the internet, messaging and notifications, we cover these cases by providing a test case for each scenario. We test the tools ability to detect the leakage going through the sink function using flow, path and context sensitive analysis 

The main sink categories are sending messages, pushing notifications and sending HTTP requests. To cover these scenarios we provide both a test where the sink is added and a test where the original sink is modified. The sinks we use are sendSms, sendPush and httpPost. They are examples where each representing one sink category respectively. The sendSms function takes a 2 parameters, a message body and a phone number, the sendPush takes 1 parameter, the notification body and the httpPost takes an array parameter which includes information necessary to make the http request such as the uri as well as the message body, then handles the response in a callback. It worth mentioning that our proposed operators address a very simplified aspect of HTTP requests in SmartThings apps, even though HTTP requests can be very complex as they may have different methods like GET, POST, PUT, DELETE, etc. and may have different content types like application/json, application/xml, or application/x-www-form-urlencoded. This is because the scope of the analysis is tainted-flows in the source code, so we don’t explore the networking side deeply. For this category of mutation, our framework injects sensitive data in the body parameter of the Groovy http request, but the functionality should be the same if the flow is injected in other parameters in the code and is passed to the http request.

\subsection{Generation of Tool-specific Reports}\label{subsec:EvaluationTools}
In this next step, the mutants are taken as input to evaluate the available static analysis tools. If the dataset is small, all the mutants will be passed to each tool and generate a tool-specific report. Alternatively, a percentage of randomly selected mutants are passed if the dataset is large. This report will contain information about how many mutants were identified, in other words, killed, by each tool and how many were not identified and are alive.  It will also provide the mutator, or the mutational operator-specific tool evaluation report.

\hlc[highlight]{Here, we demonstrate the report structure of our framework for the evaluation of the taint-flow analysis tools. It is part of a big excel file. There are more rows and column in the actual file.  We only demonstrate the framework on three categories of mutators, \textit{flow sensitivity operators}, \textit{path sensitivity operators} and \textit{context sensitivity operators}.} 

We designed our mutators to generate one change per app. For flow sensitivity, we designed six different mutators tailored to cover the most popular sinks in SmarThings apps: \emph{SMS/message}, \emph{push} message, and \emph{HTTP} sinks. For each sink type, we have two types of mutational operators: Add or Modify. 
For path sensitivity mutators, one mutator is created to add a block that can check for path sensitivity analysis. This mutator has different versions depending on different sinks. The other three operators add tainted flows to different paths of the existing code branches. Each operator is specific to each sink. For context-sensitivity, we also add calls to functions with benign and malicious contexts for three different sinks.

To generate the tool-specific results, each original source app and it's corresponding mutants have to be run through each tool. If tools under evaluation produces different analysis results for each mutant when compared to original app analysis results, such mutant will be declared as killed, and  we put 'K' in the report for that mutant. If the analysis results on a mutant is not different from analysis results on original code ,i.e., the mutant was not detected, the status of that mutant will be live and we put 'L' in such cases. If a mutant was falsely identified as malicious, we put 'FP' in the report to denote a false positive. If a benign equivalent mutant was classified as benign we put 'TN' in the report, which stands for a true negative. Based on that, we created our results for each tool regarding each mutational operator. Table \ref{fig:ReportStructure} shows a portion of the generated report for revealing the structure of the report.

\begin{table}[!ht]
\caption{Report Structure}
\centering
\resizebox{\linewidth}{!}{%
\begin{tabular}{|l|c|c|c|c|c|} 
\hline
\multicolumn{1}{|c|}{\textbf{File Name }} & \begin{tabular}[c]{@{}c@{}}\textbf{Mutation ADD }\\\textbf{possible? }\end{tabular} & \begin{tabular}[c]{@{}c@{}}\textbf{MSGFlow1 }\\\textbf{(TaintTings) }\end{tabular} & \begin{tabular}[c]{@{}c@{}}\textbf{MSGFlow1 }\\\textbf{(SAINT) }\end{tabular} & \begin{tabular}[c]{@{}c@{}}\textbf{MSGFlow2 }\\\textbf{(TaintThings) }\end{tabular} & \begin{tabular}[c]{@{}c@{}}\textbf{MSGFlow2 }\\\textbf{(SAINT) }\end{tabular}  \\ 
\hline
BeaconThingsManager                       & not possible                                                                           & N/A                                                                                & N/A                                                                           & N/A                                                                                 & N/A                                                                            \\ 
\hline
camera-motion                             & yes                                                                                & FP                                                                                 & TN                                                                            & K                                                                                   & K                                                                              \\ 
\hline
Color Temp via Virtual Dimmer             & yes                                                                                & FP                                                                                 & TN                                                                            & K                                                                                   & K                                                                              \\ 
\hline
ecobeeGetTips                             & yes                                                                                & FP                                                                                 & TN                                                                            & K                                                                                   & K                                                                              \\ 
\hline
garage-switch                             & yes                                                                                & FP                                                                                 & TN                                                                            & K                                                                                   & K                                                                              \\ 
\hline
Hue Party Mode                            & yes                                                                                & FP                                                                                 & TN                                                                            & K                                                                                   & K                                                                              \\ 
\hline
hvac-auto-off.smartapp                    & yes                                                                                & FP                                                                                 & TN                                                                            & K                                                                                   & K                                                                              \\ 
\hline
JSON                                      & yes                                                                                & FP                                                                                 & TN                                                                            & K                                                                                   & K                                                                              \\ 
\hline
Light-Alert-SmartApp                      & yes                                                                                & FP                                                                                 & TN                                                                            & K                                                                                   & K                                                                              \\ 
\hline
Lutron-Group-Control                      & yes                                                                                & FP                                                                                 & TN                                                                            & K                                                                                   & K                                                                              \\ 
\hline
Switches                                  & yes                                                                                & FP                                                                                 & TN                                                                            & K                                                                                   & K                                                                              \\ 
\hline
VirtualButtons                            & yes                                                                                & FP                                                                                 & TN                                                                            & K                                                                                   & K                                                                              \\
\hline
\end{tabular}
}
\label{fig:ReportStructure}
\caption*{FP = False Positive, TN = True Negative, K = Killed}

\end{table}
\subsection{Statistical Analysis}\label{subsec:StatisticalAnalysischap4}
In the final step, the statistical data analysis and reporting are completed for each static analysis tool. This module takes the mutants' meta-data repository from \ref{subsec:autoMutation} and the mutational tool evaluation report from \ref{subsec:EvaluationTools} as input. From this input, it computes the precision and recall that can give us an overview of the performance of the tool. 

The methodology is covered here briefly. In the last subsection, we described how the report structure of the tools is organised, and here we focus on how use the data in that report. 
Below, we present the equations that we use to compute the recall and precision.
    
\subsubsection{Measurement of Recall}\label{sec:Measurement of recall}
We need to measure recall in order to quantitatively evaluate the security analysis tools.  The proposed framework will be able to report recall for each mutator, each sensitivity type, and each static analysis tool. We follow the information retrieval's definition of recall\cite{manning2010introduction}, where the recall is the number of mutants that are detected divided by the total number of all mutants. \hlc[highlight]{Using the standard definition of recall, it is the rate of successfully identified relevant data, or true positives (TP) out of all the relevant data, the true positives and false negatives (TP + FN). }

\[Recall = \frac{TP}{TP + FN} \]
    
\hlc[highlight]{When testing for the ability of detecting the change in each mutant from the original app, in other words, if the mutant is killed by the tool being tested, the recall could also be expressed as:} 

\[ Recall = \big\{^{1,\: if \:change \:in \:the \:mutant \:in \:respect \:to \:the \:original \:app \:is  \:detected \:by \:T }_{0,\:otherwise}\]

\subsubsection{Measurement of precision}\label{sec:Measurement of precision}
In information retrieval, precision measures the noise in the results, such as irrelevant items appearing in the results of a query \cite{kontogiannis1997evaluation}. The relationship between recall and precision is an indication of how well a matching engine performs \cite{roy2009mutation}. Ideally, precision should remain high as recall increases, but in practice, this is difficult to achieve \cite{roy2009mutation}. \hlc[highlight]{Using the standard definition of precision, it is the rate of relevant identified data, or true positives (TP) out of all the identified data, the true positives and false positives (TP + FP). }

\[Precision = \frac{TP}{TP + FP} \]

\hlc[highlight]{This can also be expressed in terms of how many of the killed mutants \textit{k} are correctly identified to contain relevant data that is valid \textit{v}. So the precision can be calculated as follows:}
\[ Precision =\frac{v}{k}\]

\hlc[highlight]{These equations can be applied per mutator or for one category of mutators addressing specific sensitivity (flow, path, or context) analysis and that by averaging the calculated precision and recall of the mutators in each category. Similarly, to calculate the overall precision and recall for a specific tool, we give equal weight to each of the three categories of sensitivity analysis mutators, so we can average their results.}


\section{Experiment}\label{taset}
\hlc[highlight]{For the experiment, we generated mutants using our framework from a dataset of SmartThings applications. We then run the static taint analysis tools on the generated mutants and the original source apps. We then compare their results to see whether they successfully identified the mutants and whether it was correctly marked as malicous or benign.} 

\subsection{Dataset}
 We used a dataset comprised of different kinds of SmartThings applications including IoTBench, an open repository for evaluating systems designed for IoT app analyses \cite{IoTBench-test-suite}. The dataset is the one used by Celik et al. \cite{celik2018sensitive} for SaINT and Schmeidl et al. \cite{schemild} for Taint-Things. The dataset includes 264 applications; 42 official SmartThings Marketplace apps, 144 official apps provided by the community, 59 third-party apps, and 19 apps specifically developed by the SaINT team to include common vulnerabilities available online under IoTBench test suite \cite{schemild}. 
 
 \hlc[highlight]{Our framework was developed before Taint-things and FlowsMiner were available. SAINT was the only available tool, so we used it to initially identify the benign dataset and we then manually checked the dataset to confirm its benign status. As the other tools,  Taint-Things and FlowsMiner, became available, and to guarantee that the dataset is not biased, we selected benign applications which were confirmed benign using these tools as well.} We picked the \hlc[highlight]{apps} that were marked by the three tools as benign with the exception of one \hlc[highlight]{app}  that Taint-Things marked as tainted that was used to generate the context sensitivity tests. We have included it due \hlc[highlight]{to} the limitation of original \hlc[highlight]{apps} fit for context sensitivity. We address this in more detail in Section \ref{ch: context -Automated Mutants Generation}.

\begin{figure}[!htb]
	\centerline{\includegraphics[width=.99\textwidth]{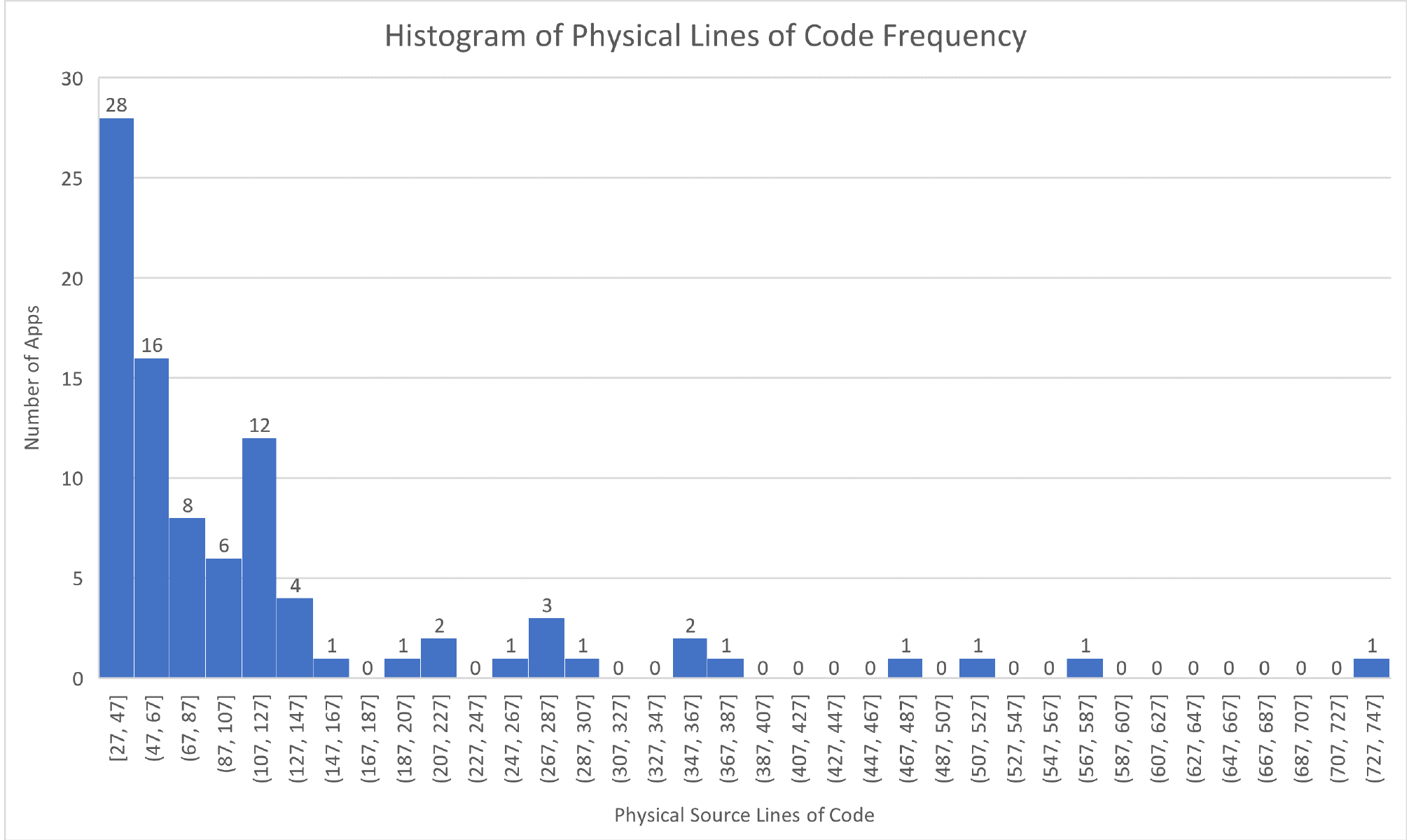}}
	\caption{Histogram of Physical Lines of Code Frequency}
	\label{fig:LOC}
\end{figure}

As a result, we got 65 community apps, 13 forum apps, 11 marketplace apps, and one app from SaINT IoTbench. The distribution of the different source \hlc[highlight]{apps} and \hlc[highlight]{mutants} created for each mutational operator in the dataset is given in Table \ref{tab:Files created for each mutational operators}. \hlc[highlight]{Each app is a single file and range from 27 to 732 physical lines of code (LOC), with the blank and comment lines excluded. The average LOC is 118 and the median is 65.5. Figure} \ref{fig:LOC} \hlc[highlight]{shows a histogram of the apps' LOC. Generally the SmartThings apps are contained in a single-file, furthermore, the scope of the research and the tested tools analyze single-file apps.} 

\begin{table}[ht]
\centering
\caption{\label{tab:Files created for each mutational operators}\hlc[highlight]{Apps and mutants}  created for each mutational operator}
\begin{tabular}{|l|l|l|l|l|}
\hline
                 & SaINT & Forum & Marketplace & Community \\ \hline
Apps in Dataset & 1     & 13    & 11          & 65        \\ \hline
MMfs             & 0     & 0     & 0           & 0         \\ \hline
MPfs             & 0     & 0     & 0           & 0         \\ \hline
MHfs             & 0     & 0     & 0           & 0         \\ \hline
AMfs             & 2     & 24    & 22          & 126       \\ \hline
APfs             & 2     & 24    & 22          & 126       \\ \hline
AHfs             & 2     & 24    & 22          & 126       \\ \hline
Aps              & 6     & 72    & 66          & 378       \\ \hline
AMps             & 8     & 56    & 62          & 330       \\ \hline
APps             & 8     & 56    & 62          & 330       \\ \hline
AHps             & 8     & 56    & 62          & 330       \\ \hline
AMcs             & 0     & 0     & 6           & 26        \\ \hline
APcs             & 0     & 0     & 6           & 26        \\ \hline
AHcs             & 0     & 0     & 6           & 26        \\ \hline
\end{tabular}
\end{table}

\subsection{Tested Tools}
We are aiming to evaluate static analysis tools that use the taint flow analysis technique and are designed to ensure the security of the IoT SmartThings platform. Currently, we only found three such tools in this developing field. They are SaINT \cite{celik2018sensitive}, Taint-Things \cite{schemild}, and FlowsMiner \cite{DBLP:conf/wcre/NaeemA20}.

\textbf{SaINT:} The first tool to be found that matches the criteria we were looking into is SaINT, a tool that is designed to analyze sensitive data flow and leakage in IoT applications. It uses static analysis techniques.
SaINT operates in three phases:  Translation of platform-specific IoT source
code into an intermediate representation (IR), then identifying sensitive sources and sinks, and finally, performing static analysis to identify sensitive data flows \cite{celik2018sensitive}. The IR is constructed from the source code of the inputted IoT app and it requires building the app’s inter-procedural control flow graph (ICFG). SaINT’s IR-building algorithm directly works on the Abstract Syntax Tree (AST) representation of Groovy code \cite{celik2018sensitive}. SaINT identifies flows from source to sink. The tool is available at \cite{SAINT}.

\textbf{Taint-Things:} Taint-Things provides a static analysis approach to identify and report tainted flows in SmartThings Apps \cite{schemild}. Taint-Things is built by using the TXL language, which needs a grammar with the capability to define the language of the inputted text. The authors used the same Groovy grammar as us for their implementation.  Taint-Things does backward tracing for its flow from the sinks to the sources. For addressing a different kind of sensitivity it has different versions. For flow sensitivity, it has two versions published on its website: one including Static Single Assignment (SSA) and one without SSA.  Path and context-sensitivity specific versions were provided as options for the analysis and the tool is available publicly \hlc[highlight]{online} \cite{TaintThings}. 
We checked different kinds of sensitivity operators in Taint-Thing's different versions.

\textbf{FlowsMiner:} FlowsMiner uses text mining techniques to identify tainted flows. It transforms the source code into a standard format before tokenizing it. It starts by identifying the sources of potential sensitive data, then it reads the potential leakage sinks. This tool uses a fixed set of sinks provided in a text file and provides flexibility to update it and for the user to provide his own set. The tool tokenizes the source code to check if it contains a sink. If the source code contains a sink, it proceeds further to identify tainted flows. The tool is available \hlc[highlight]{online} \cite{FlowsMiner}. 
  

\section{Flow-sensitive Mutation Generation and Analysis}\label{ch: Flow -Automated Mutants Generation}
Our framework proposes six new mutational operators for flow-sensitivity.  We exploited three commonly used sinks in the field of IoT SmartThings to create these flow-sensitive operators. We either proposed to add or modify the content of an existing program. We have three add and three modify operator for flow. Table \ref{tab:flow mutators} outlines the proposed set of mutational operators for flow-sensitivity. The first column represents the name of the operator, the second column presents a brief description of the operator and the last column indicates the type of sensitivity analysis that the mutational operator belongs to.
\begin{table*}[!b]
\centering
\caption{\label{tab:flow mutators} Proposed flow- sensitive mutational operators}
\begin{tabular}{|l|l|l|l|} 
\hline
\textbf{ Name}  & \textbf{~ ~ ~ ~ ~ ~ ~ ~ ~ ~ ~ ~ ~ ~ ~ ~ ~ ~ ~Description }                                                                                                                                 & \textbf{Type }  \\ 
\hline
\textit{MMfs }  & Modify the information sent in a message                                                                                                                                                                    & Flow            \\ 
\hline
\textit{AMfs }  & Add a send message sink with sensitive information                                                                                                                                                         & Flow            \\ 
\hline
\textit{MPfs }  & Modify the information sent in a Push message                                                                                                                                                               & Flow            \\ 
\hline
\textit{APfs }  & Add a Push message sink with sensitive information                                                                                                                                                         & Flow            \\ 
\hline
\textit{MHfs }  & Modify the information(body) sent in a HTTP Post                                                                                                                                                            & Flow            \\ 
\hline
\textit{AHfs }  & Add a HTTP sink with sensitive information                                                                                                                                                                 & Flow            \\ 
\hline
\end{tabular}
\end{table*}
\subsection{Flow-sensitive Mutants Generation}\label{sec:How the flow-sensitive Mutants are generated}
To automate the mutant generation stage, we use TXL source transformation technology \cite{cordy2009excerpts}. The transformation process takes SmartThings apps as input, which are written in Groovy, and hence, a TXL Groovy grammar is also required. We used a TXL Groovy grammar developed by Schmeidl et al. \cite{schemild}. We redefined the grammar when required.  Those apps need to be parsed and analyzed to enable automated mutation injection. Based on the set of mutational operators we proposed in this study and listed in Table \ref{tab:flow mutators}, we have developed several mutators and TXL transformation rules to automatically generate the mutants. Each mutator will generate one mutation, meaning only one fault is injected in each SmartThings app. 

This stage also generates a mutant meta-data repository and enables tracking of how many mutants were caught and killed by static analysis tools used in the evaluation process. This meta-data repository will store the information related to flow-sensitive mutants. Listings \ref{list:preference flow}-\ref{list:Mutated function1} demonstrate the mutational injection process on a SmartThings app. The mutator used in this example is \textit{MMfs}, Modify Message flow sensitivity.

To begin, the mutator will pattern match the preference block in the SmartThings App, and find and extract one sensitive source as defined in the input section. This is shown in Listing \ref{list:preference flow}.

\begin{lstlisting}[style=Groovy, caption={Preference}, label=list:preference flow]
preferences {
  section("When all of these people leave home") {
    input ~"people"~, "capability.presenceSensor", multiple: true }
}
\end{lstlisting}

Afterwards, it will pattern match a function that has a sink of type \textit{sendSMS} and extract the variable name that the sink uses to pass information. In this example, the variable name is \textit{messages} as shown in Listing \ref{list:function}. 
 
\begin{lstlisting}[style=Groovy, caption={Function}, label=list:function]
def eventHandler(evt) { 
  def messages = "new event"
  ~sendSMS("1111111111", messages)~ 
}
\end{lstlisting}

The \textit{MMfs} mutator will then construct the injection vector, in this case, it is an assignment statement that has variable messages in the left hand side and the sensitive information \textit{\$people} in the right hand side. The mutation vector is then injected just before the sink, \textit{sendSMS()} as shown in Listing \ref{list:Mutated function1}.

\begin{lstlisting}[style=Groovy, caption={Mutated function}, label=list:Mutated function1]
def eventHandler(evt) { 
  def messages = "new event"
  `messages = "${people}"`
  sendSMS("1111111111", messages)
}
\end{lstlisting}
Knowing how to generate flow-sensitive mutants, we can understand the description of the mutational operators better. In the next subsections, we are going to describe each flow-sensitive mutator and their output for SaINT, FlowsMiner and Taint-Things.

\subsubsection{Modify Message Flow-sensitivity (MMfs)}
This operator modifies the string that is supposed to be sent by a \textit{sendSms} sink to send a vulnerable string instead. The mutator constructs the vulnerable string by appending an input variable from the preferences section of the SmartThings app. 

\begin{figure*}[!ht]
	\centerline{\includegraphics[width=.99\textwidth]{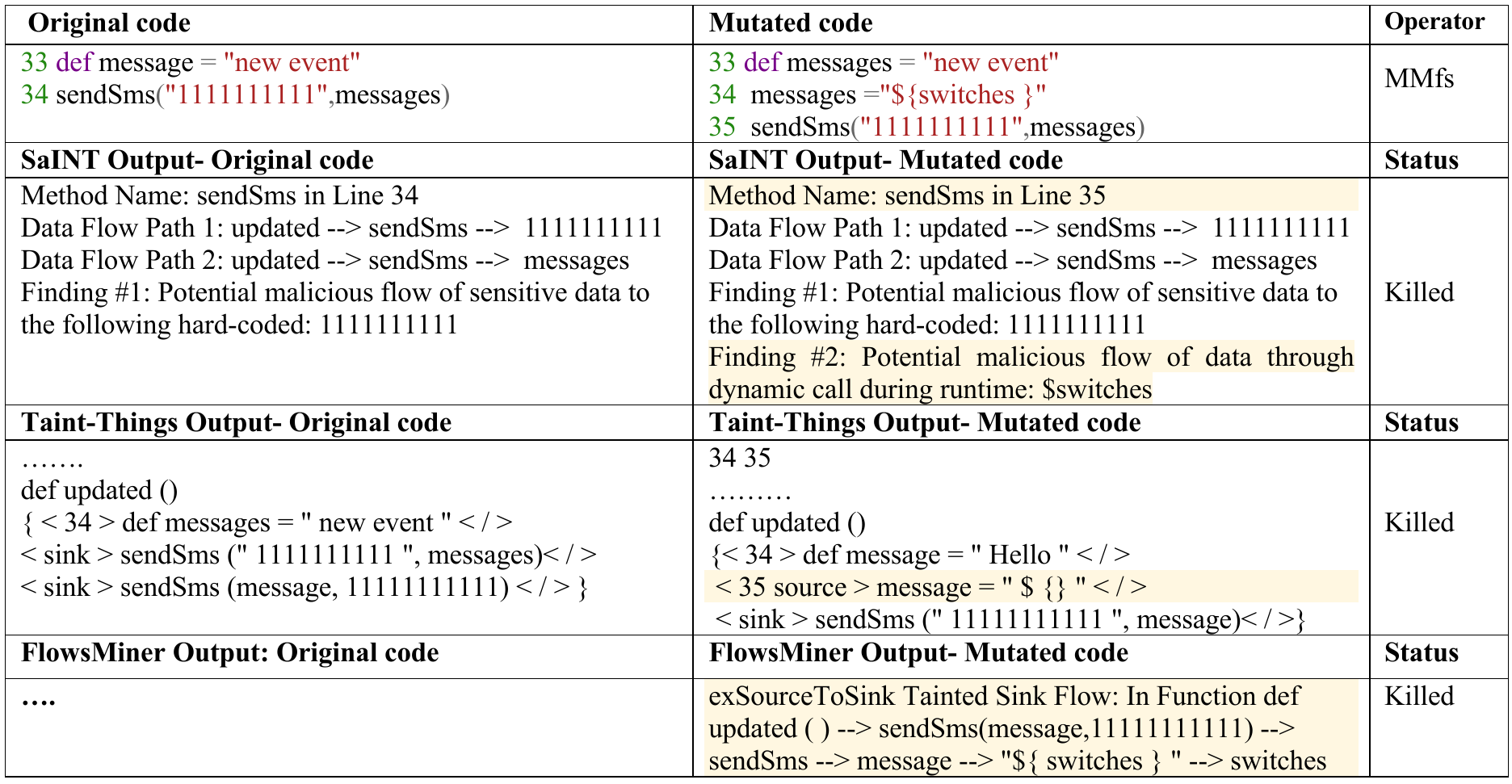}}
	\caption{Modify Message Flow-sensitivity (MMfs) Operator}
	
	\label{fig:MMfs}
\end{figure*}

Figure \ref{fig:MMfs} presents an example where the mutator changed the value of the \textit{message} variable  before sending it through \textit{sendSms()} sink. The change appends an input variable \textit{\$switch} to the message, thus making it vulnerable. SaINT, Taint-things and FlowsMiner were able to kill that mutant. 

\subsubsection{Modify Push-message Flow-sensitivity (MPfs)}
This operator works in a similar way to MMfs. The only difference is that the MPfs will change the string before sending it to a \emph{push} notification instead of \emph{SMS}.
 
\begin{figure}[!t]
	\centerline{\includegraphics[width=.99\textwidth]{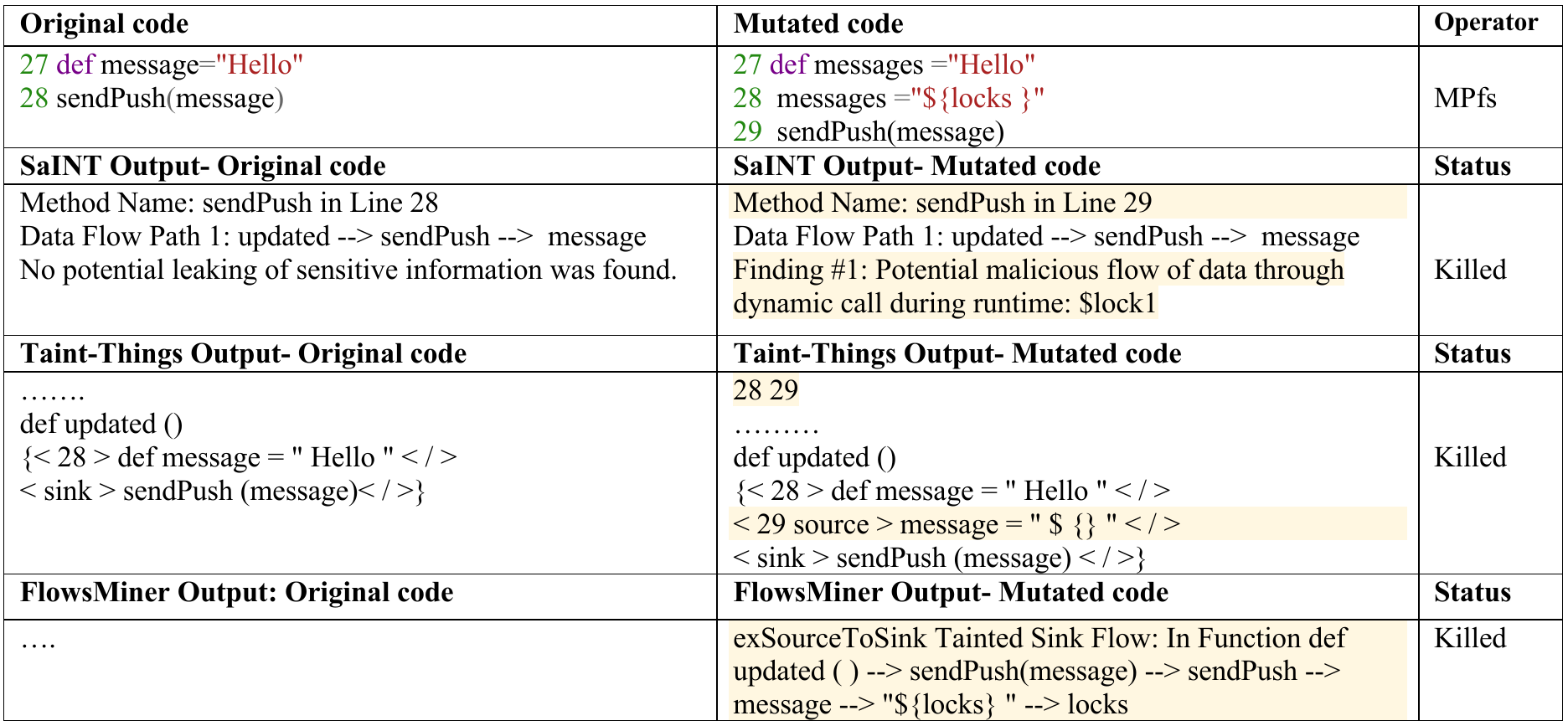}}
	\caption{Modify Push-message Flow-sensitivity (MPfs) Operator}

	\label{fig:MPfs}
\end{figure}

Figure \ref{fig:MPfs} demonstrates a mutator that changes the value to be sent by \textit{sendPush} message from a non-vulnerable string, \textit{"Hello"} to a vulnerable string, leaking an inputted variable, \textit{\$lock}. Both tools were able to kill the mutant.

\subsubsection{Modify HTTP-Request Flow-sensitivity (MHfs)}
\begin{figure}[!t]
	\centerline{\includegraphics[width=.99\textwidth]{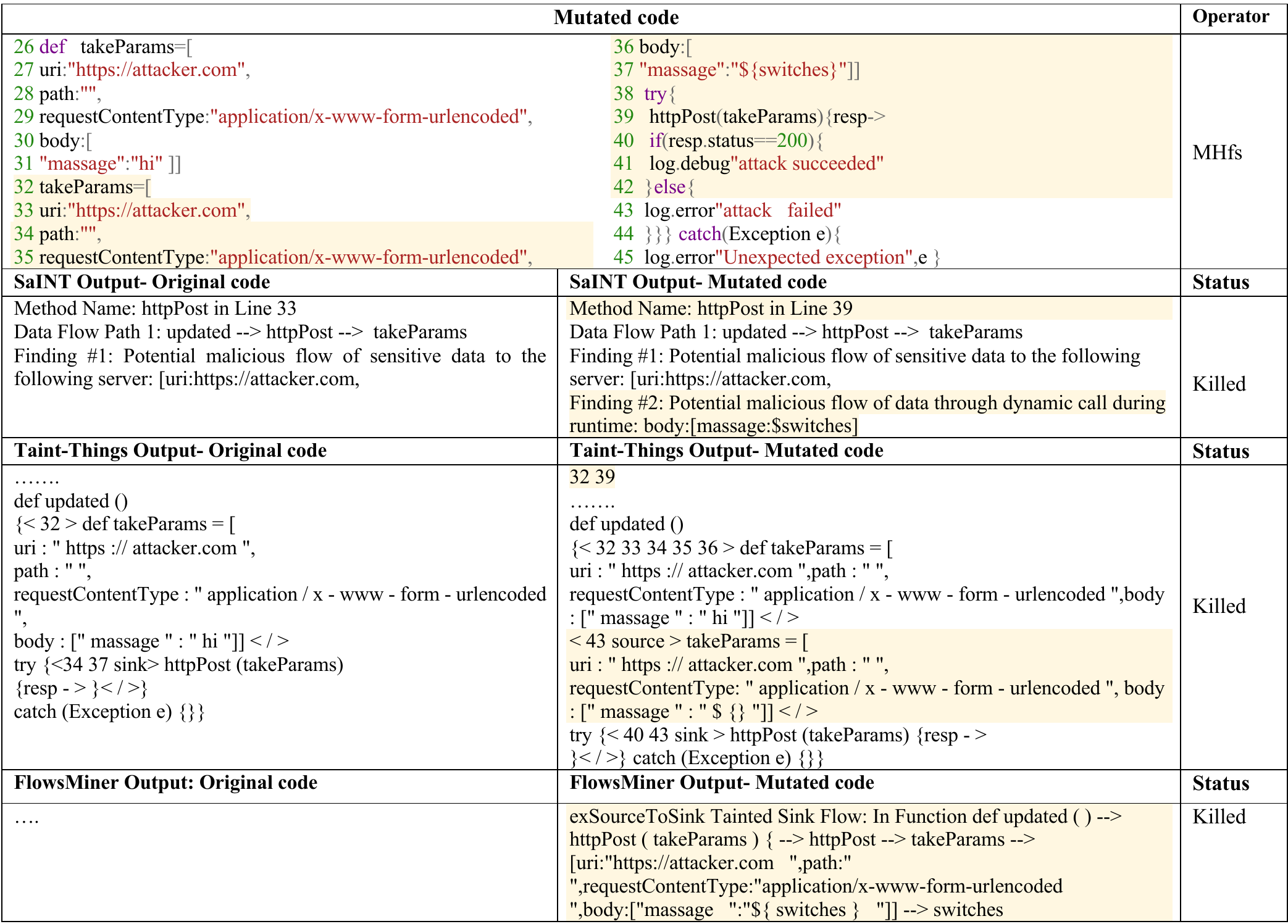}}
	\caption{Modify HTTP-Request Flow-sensitivity (MHfs) Operator}
	
	\label{fig:MHfs}
\end{figure}

This operator exploits the content sent by an internet protocol. If there is any post method called to post something to a website, it will replace the body of the content with a vulnerable input. In Figure \ref{fig:MHfs}, we only present the mutant with the highlighted injected code; the non-highlighted lines are included in the original code. Both tools were able to kill the mutant. 

\subsection{Add Message Flow-sensitivity (AMfs)}\label{subsec:AMFS}
AMfs is one of the sets of \textit{Add} mutational operators where a complete flow vector is added to the original file. This includes a sink of type \textit{sendSms}, a benign message to create a file with a benign flow, or a vulnerable message to create a file with a vulnerable flow.  

\begin{figure}[!t]
	\centerline{\includegraphics[width=.99\textwidth]{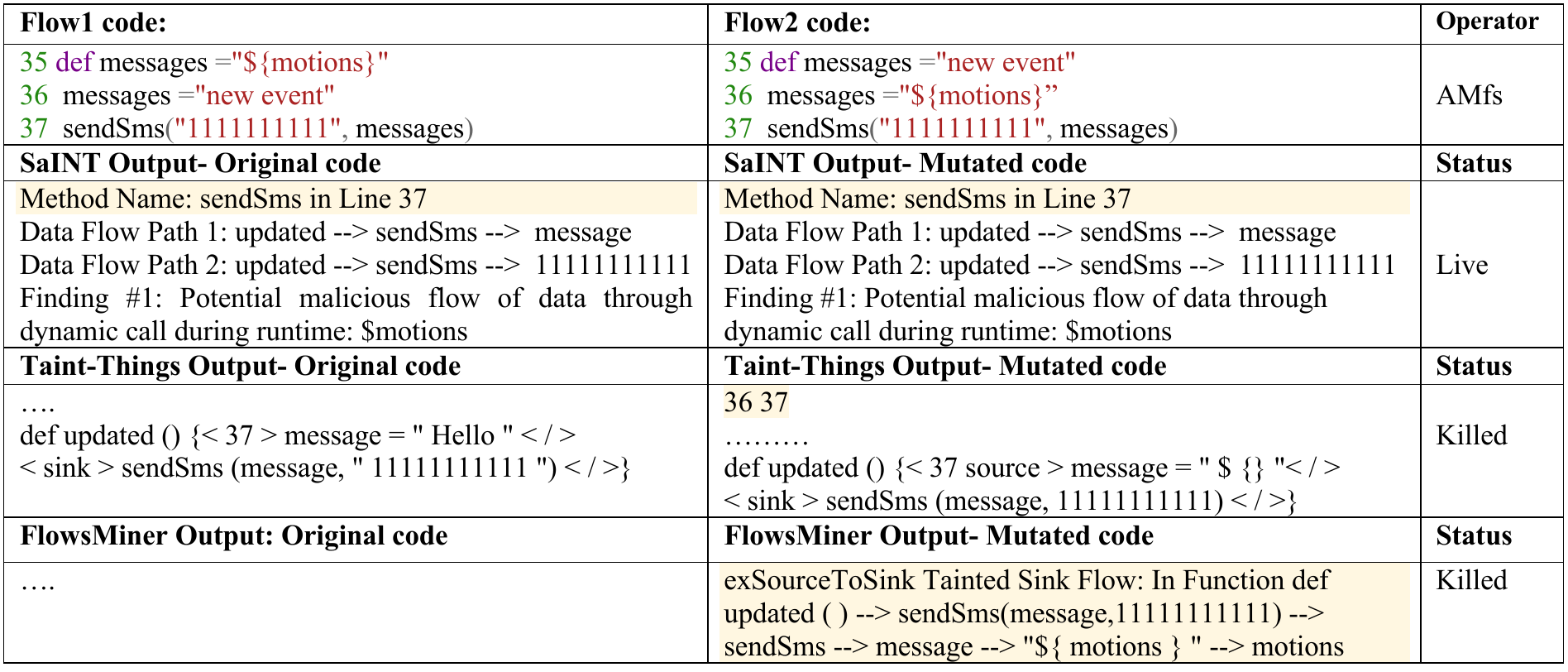}}
	
	\caption{Add Message Flow-sensitivity (AMfs) Operator}
	
	\label{fig:AMfs}
\end{figure}

The goal is to evaluate the tools' flow-sensitive analysis so that a benign file is identified as a benign and vulnerable file is identified as such. A flow-insensitive analysis will fail to make such a distinction. %
Figure \ref{fig:AMfs} presents an example in which the first flow, \textit{Flow1}, the mutator constructs a string with a vulnerable input value \textit{\$motion}, which will be replaced with a harmless value afterwards. Then, the string will be sent out through a \emph{sendSms} sink. It's a benign flow, as the value sent is not vulnerable. In \textit{Flow2}, the mutator constructs the two assignment statements in the opposite order, the first value to put in the string is a harmless one. Then, after replacing it with vulnerable input value, it will be sent through the \textit{sendSms} sink. As such, \textit{Flow2} is vulnerable. The value of the input variable \textit{\$motion} will be replaced by a fixed string at line 36, which makes the flow benign if the analysis is flow-sensitive. Taint-Things is a flow-sensitive analyzer that kills those mutants precisely. However, SaINT fails to identify/kill them if we consider flow1 as base code and flow2 as mutated code. SaINT identifies both flows as malicious, where Taint-Things and FlowsMiner have the ability to distinguish the benign flow from the vulnerable one. The benign flow, \textit{Flow1}, was reported vulnerable by the SaINT tool, and that is false positive. 

\subsubsection{Add Push-message Flow-sensitivity (APfs)}\label{subsec:APFS}
\begin{figure}[!t]
	\centerline{\includegraphics[width=.99\textwidth]{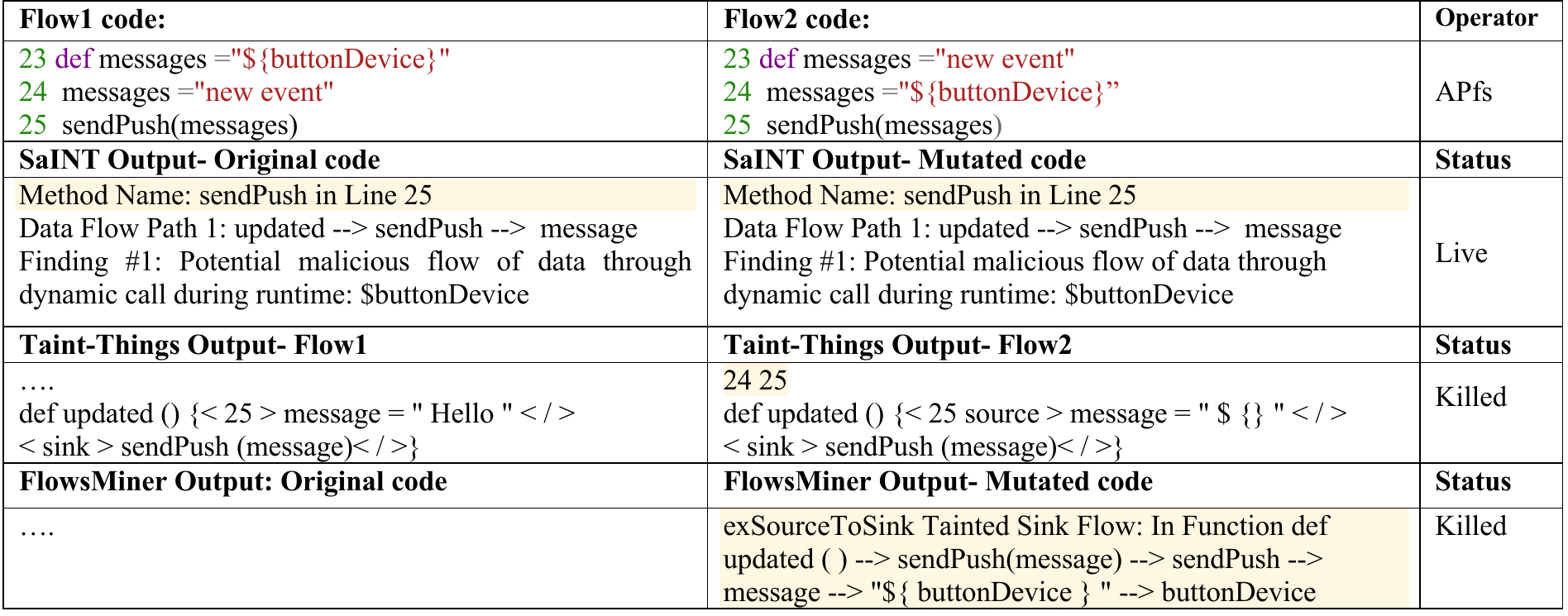}}
	\caption{Add Push-message Flow-sensitivity (APfs) Operator}
	\label{fig:APfs}
\end{figure}

This operator is structured in the same way as AMfs. Here, instead of constructing a \emph{sendSms} sink, the information is sent by \emph{push}
notification. The mutator also constructs two mutants, one benign and the other,
vulnerable. Figure \ref{fig:APfs} shows an example of two flows. \textit{Flow1} starts with an input variable \textit{\$buttonDevice}, which is then replaced by a harmless string value. As such, Flow1 is benign. However, \textit{Flow2} is vulnerable.  Taint-Things and and FlowsMiner identify both flows precisely whereas SaINT does not and ends up creating false-positive results for the benign one.

\subsubsection{Add HTTP-Request Flow-sensitivity (AHfs)}\label{subsec:AHFS}
\begin{figure*}[!ht]
	\centerline{\includegraphics[width=.99\textwidth]{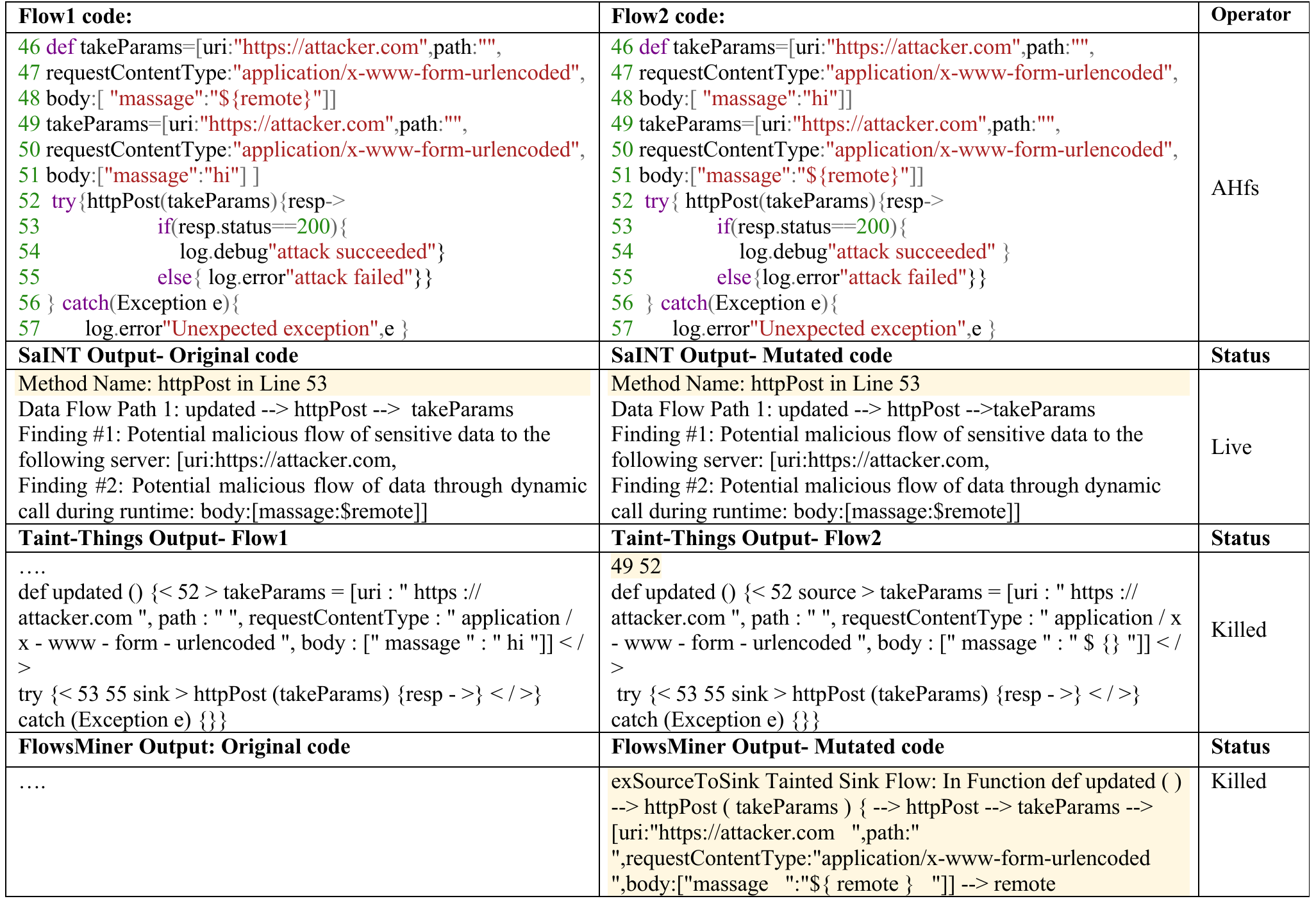}}
	\caption{Add HTTP-Request Flow-sensitivity (AHfs) Operator}
	
	\label{fig:AHfs}
\end{figure*}

This operator is designed to add content with flow sensitivity sent by an internet protocol. It adds a \textit{httpPost} method for a website with a vulnerable input. Like all other add operators, we generate two types of flows for this operator, one benign and one vulnerable. SaINT detects both flows as vulnerable, where \textit{Flow1} gets identified as a false-positive. Taint-Things and FlowsMiner are able to identify benign and vulnerable flows precisely. Figure \ref{fig:AHfs} shows two different flows where \textit{httpPost} is called with two different values. However, SaINT fails to identify the difference. For \textit{Flow1}, the function will be called with no input variable in it as it gets replaced in the flow. Nevertheless, SaINT still identifies it as vulnerable as SaINT does the flow-insensitive analysis.

\subsection{Flow-sensitive Mutants' Experiment and Results}\label{sec:Flow Sensitive mutants' experiment and Result}
 
\begin{table*}[!b]
\centering
\caption{\label{tab:Mutants executed}Files executed for flow-sensitivity in different security tools}
\begin{adjustbox}{width=0.99\textwidth,totalheight={0.99\textheight},keepaspectratio}%
\begin{tabular}{lllllllllllll}
\hline
Tool         & MMfs                                                       &                                                         & AMfs                                                       &                                                         & MPfs                                                       &                                                         & APfs                                                       &                                                         & MHfs                                                           &                                                         & AHfs                                                       &                                                         \\ \cline{2-13} 
             & \begin{tabular}[c]{@{}l@{}}Files\\ Generated\end{tabular} & \begin{tabular}[c]{@{}l@{}}Files\\ 
             Executed\end{tabular} & \begin{tabular}[c]{@{}l@{}}Files\\ Generated\end{tabular} & \begin{tabular}[c]{@{}l@{}}Files\\ 
             Executed\end{tabular} & \begin{tabular}[c]{@{}l@{}}Files\\ Generated\end{tabular} & \begin{tabular}[c]{@{}l@{}}Files\\ Executed\end{tabular} & \begin{tabular}[c]{@{}l@{}}Files\\ Generated\end{tabular} & \begin{tabular}[c]{@{}l@{}}Files\\ Executed\end{tabular} & \begin{tabular}[c]{@{}l@{}}Files\\ Generated\end{tabular} & \begin{tabular}[c]{@{}l@{}}Files\\ Executed\end{tabular} & \begin{tabular}[c]{@{}l@{}}Files\\ Generated\end{tabular} & \begin{tabular}[c]{@{}l@{}}Files\\ Executed\end{tabular} \\ \hline
SaINT        & 0                                                          & 0                                                       & 174                                                         & 162                                                      & 0                                                          & 0                                                       & 174                                                         & 164                                                      & 0                                                          & 0                                                       & 174                                                         & 170                                                      \\
Taint-Things & 0                                                          & 0                                                       & 174                                                         & 174                                                      & 0                                                          & 0                                                       & 174                                                         & 174                                                      & 0                                                          & 0                                                       & 174                                                         & 174                                                     \\
FlowsMiner & 0                                                          & 0                                                       & 174                                                         & 170                                                      & 0                                                          & 0                                                       & 174                                                         & 170                                                      & 0                                                          & 0                                                       & 174                                                         & 170                                                     \\
\hline
\end{tabular}
\end{adjustbox}

\end{table*}

\begin{table}[t!]
\centering
\caption{\label{tab:Mutants Killed flow sensitive}Mutants result for flow sensitivity}
\resizebox{\linewidth}{!}{%
\begin{tabular}{|l|l|l|l|l|l|l|l|l|l|} 
\hline
\begin{tabular}[c]{@{}l@{}}Mutator\\ (2files \\ creating\\ one \\ mutation) \end{tabular} & Mutant              & \begin{tabular}[c]{@{}l@{}}SaINT\\ Result \end{tabular} & \begin{tabular}[c]{@{}l@{}}Taint-\\ Things\\ Result \end{tabular} & \begin{tabular}[c]{@{}l@{}}Flows-\\Miner\\Results\end{tabular} & \begin{tabular}[c]{@{}l@{}}Mutator \\ version\\ (Each file one \\ mutant) \end{tabular} & Mutant & \begin{tabular}[c]{@{}l@{}}SaINT\\ Result \end{tabular} & \begin{tabular}[c]{@{}l@{}}Taint-\\ Things\\ Result \end{tabular} & \begin{tabular}[c]{@{}l@{}}Flows-\\Miner\\Results\end{tabular}  \\ 
\hline
\multirow{2}{*}{AMfs}                                                                     & \multirow{2}{*}{87} & \multirow{2}{*}{81L}                                    & \multirow{2}{*}{87K}                                              & \multirow{2}{*}{85K}                                          & AMfsflow1                                                                               & 87     & 81FP                                                    & 87TN                                                              & 85TN                                                           \\ 
\cline{6-10}
                                                                                          &                     &                                                         &                                                                   &                                                               & AMfsflow2                                                                               & 87     & 81K                                                     & 87K                                                               & 85K                                                            \\ 
\hline
\multirow{2}{*}{APps}                                                                     & \multirow{2}{*}{87} & \multirow{2}{*}{82L}                                    & \multirow{2}{*}{87K}                                              & \multirow{2}{*}{85K}                                          & APfsflow1                                                                               & 87     & 82FP                                                    & 87TN                                                              & 85TN                                                           \\ 
\cline{6-10}
                                                                                          &                     &                                                         &                                                                   &                                                               & APfsflow2                                                                               & 87     & 82K                                                     & 87K                                                               & 85K                                                            \\ 
\hline
\multirow{2}{*}{AHps}                                                                     & \multirow{2}{*}{87} & \multirow{2}{*}{85L}                                    & \multirow{2}{*}{87K}                                              & \multirow{2}{*}{85K}                                          & AHfsflow1                                                                               & 87     & 85FP                                                    & 87TN                                                              & 85TN                                                           \\ 
\cline{6-10}
                                                                                          &                     &                                                         &                                                                   &                                                               & AHfsflow2                                                                               & 87     & 85K                                                     & 87K                                                               & 85K                                                            \\
\hline
\end{tabular}
}
\caption*{K = Killed, L = Lived, FP = False Positive, TN = True Negative}
\end{table}
In this section, we demonstrate the use of the framework on evaluating tools for flow-sensitive analysis. We focus on the category of flow-sensitive mutational operators, as such, we have developed six mutators.When applying the mutators to the set of benign apps, we realized that no mutants were created for MMfs, MPfs, and MHfs. We further checked the code manually and found that there were no existing sinks in the sequential flow, so the \textit{Modify} operators did not work. Instead, all the mutants created were generated using the AMfs, APfs, and AHfs operators.

For each Add operator and from 90 benign apps we were able to generate 174 \hlc[highlight] {mutants}. From three \hlc[highlight] {apps}, no mutations were created, as there is no source or input there. For each type of Add operator, we generated two \hlc[highlight] {mutants}: one benign, which does not have any taint-flow vulnerability and is aimed at evaluating tools' false positives; and another, vulnerable, which is aimed at evaluating tools' false negatives. For the AMfs, APfs, and AHfs we were leaking the information that was defined in the input section. For three \hlc[highlight] {apps}, there were no sensitive data defined in the input section, so there was no mutant for the three types of Add mutators for those \hlc[highlight] {apps}. Next, we used our generated mutants to evaluate three existing static analysis tools designed for taint flow identification, SaINT, Taint-Things and FlowsMiner. Table \ref{tab:Mutants executed} shows the number of files that were executed by each of tools from the generated mutants. \hlc[highlight]{In our evaluation, cases where the tools times out or do no executed on certain mutants are excluded in the calculation of precision and recall, since they do not give us relevant data. This nonetheless is relevant if one is comparing the general performance of the tools.}

If we consider each file as a mutant, Taint-things was able to identify all of the generated 87 leaking mutants for each of the 3 Add operators shown in Table \ref{tab:Mutants Killed flow sensitive} while FlowsMiner is able to identify 85, since it failed to run on the mutators generated from 2 original files. In Table \ref{tab:Mutants Killed flow sensitive} in the SaINT, Taint-Things and FlowsMiner results columns the letter after the number denotes it's classification. `K' stands for `Killed', `L' stands for `Live', `FP' stands for `False positive' and `TN' stands for `True Negative'. Taint-things also identifies 87 equivalent mutants that were benign (true negative). It did not give us any false positives. Similarly, FlowsMiner identifies the 85 equivalent mutators it ran on without any false positives So the precision and recall are 100\%. SaINT was not able to finish running 12 AMfs(6 flow1 and 6 flow2), 10 APfs(5 flow1 and 5 flow2), and 4 AHfs (2 flow1 and 2 flow2) in the community set. SaINT was facing timeout due to either the big size of the code or it containing too much function calls. As it is a server-side issue not related to mutants identification, we ignored the 26 mutants when calculating precision and recall. We did not face the timeout issue with Taint-Things as we had access to the executable. Since SaINT was also reporting the equivalent benign mutants as vulnerable, its precision dropped to 50\%. It has a high false-positive rate when it comes to flow-sensitive analysis as shown in Table \ref{tab:Precision and recall}. 

\begin{table}[!ht]
\centering
\caption{\label{tab:Precision and recall}Precision and recall for different mutators and tools}

\begin{tabular}{llllllll}
\hline
{Type}             & {Mutators} & \multicolumn{2}{l}{SaINT} & \multicolumn{2}{l}{Taint-Things} & \multicolumn{2}{l}{FlowsMiner}\\ \cline{3-8} 
                                  &                           & Rec.\%        & Prec.\%       & Rec.\%            & Prec.\%  & Rec.\%            & Prec.\%           \\ \hline
{Flow-Sensitivity}           
                                  & AMfs                      & 100         & 50         & 100             & 100  & 100             & 100           \\
                                  & APfs                      & 100         & 50         & 100             & 100   & 100             & 100          \\
                                  & AHfs                      & 100         & 50         & 100             & 100    & 100             & 100   
                                  
                                  \\ \cline{2-8} 
                                                                    & Overall                      & 100         & 50         & 100             & 100    & 100             & 100   
                                  
                                  \\ \cline{2-8} 
\end{tabular}


\end{table}

For Taint-Things and FlowsMiner all mutants were discovered, while SaINT there was a good number of false positives, but none of the vulnerable mutants went undiscovered. 

Equivalent mutants are one of the biggest problems in mutation testing.  Equivalent mutants keep the program's semantics unchanged and thus cannot be detected by any test \cite{grun2009impact}. In our approach, we generate benign mutants, which are equivalent to the base code in terms of not having a vulnerability, and that to evaluate the tools' false positive rate. \hlc[highlight]{These mutants may be equivalent, in addition to being benign, so they do not} change the semantics of the original program since they don't leak information, however, we have full control of these mutants and we use them for evaluating the tools. As such, they can not be classified as problematic equivalent mutants.

When we evaluate every flow-sensitive mutator considering flow1 as the original base-code and flow2 as a mutation, a total of 87 mutants were created, each consisting of flow1 as a base app and flow2 as a mutant. In this way, we checked if the tools can classify them correctly while addressing the flow change from benign to malicious. As we stated before, flow1 has a benign flow, while flow2 has a malicious one.

We calculated the values in Table \ref{tab:Precision and recall for 2 files} by counting flow1 as the base and flow2 as the mutated program and that to analyze flow-sensitivity accurately.  This table shows us clearly that SaINT is using a flow-insensitive analysis to detect vulnerability where Taint-Things and FlowsMiner are using a flow-sensitive analysis. For Taint-Things and FlowsMiner, the Recall and Precision rate is still 100\%, where for SaINT, it will become 0\% for both, as all the mutants will be alive in this case. They are classified as alive because SaINT can not distinguish between different flows and gives us the same output for flow1 and flow2 for each mutator, when actually one is benign and the other is malicious.

\begin{table}[!ht]
\centering
\caption{\label{tab:Precision and recall for 2 files}Precision and recall for different mutators (2 flows counted as one mutant) for SaINT and Taint-Things}

\begin{tabular}{llllllll}
\hline
{Type}             & {Mutators} & \multicolumn{2}{l}{SaINT} & \multicolumn{2}{l}{Taint-Things} & \multicolumn{2}{l}{FlowsMiner}  \\ \cline{3-8} 
                                  &                           & Rec.\%        & Prec.\%       & Rec.\%            & Prec.\%   & Rec.\%            & Prec.\%         \\ \hline
{Flow-Sensitivity}           
                                  & AMfs                      & 00         & 00         & 100             & 100  & 100             & 100           \\
                                  & APfs                      & 00         & 00         & 100             & 100    & 100             & 100          \\
                                  & AHfs                      & 00         & 00         & 100             & 100    & 100             & 100   
                                  
                                  \\ \cline{2-8} 
                                  
                                                                  & Overall                      & 00         & 00         & 100             & 100    & 100             & 100   
                                                                \\ \cline{2-8} 
    
\end{tabular}

\end{table}


\section{Path-sensitive Mutation Generation and Analysis}\label{Path Automated Mutants Generation}
Our framework presents four new mutational operators for path-sensitivity. Similar to flow-sensitive mutational operators, we used three commonly used sinks in the field of IoT SmartThings to design these path-sensitive operators. 
Table \ref{tab:Path mutators} outlines the proposed set of mutational operators for path-sensitivity  
  
\begin{table*}[!ht]
\centering
\caption{\label{tab:Path mutators} Proposed  path- sensitive mutational operators}
\begin{tabular}{|l|l|l|l|} 
\hline
\textbf{ Name}  & \textbf{~ ~ ~ ~ ~ ~ ~ ~ ~ ~ ~ ~ ~ ~ ~ ~ ~ ~ ~Description }                                                                                                                                 & \textbf{Type }  \\ 
\hline
\textit{Aps }   &\begin{tabular}[c]{@{}l@{}} Add a conditional statement with one branch \\leaking sensitive input \end{tabular}                                                                                                                                        & Path            \\ 
\hline
\textit{AMps }  & \begin{tabular}[c]{@{}l@{}}Add a send message sink to leak information if multiple paths\\ are found where~at least one path contains sensitive information \end{tabular}                                     & Path            \\ 
\hline
\textit{APps }  & \begin{tabular}[c]{@{}l@{}}Add Push sink to leak information if multiple paths are found \\where at least one path contains sensitive information \end{tabular}                                               & Path            \\ 
\hline
\textit{AHps }  & \begin{tabular}[c]{@{}l@{}}Add HTTP sink to leak information if multiple path is found \\where at least one path contains sensitive information \end{tabular}                                               & Path            \\ 

\hline

\end{tabular}
\end{table*}
\subsection{Path-sensitive Mutants Generation}\label{sec:How the path-sensitive Mutants are generated}
During the mutant generation stage for path-sensitivity, we use the same TXL source transformation technology \cite{cordy2009excerpts} and the same TXL Groovy grammar \cite{schemild} as the flow-sensitive analysis mentioned in Section \ref{ch: Flow -Automated Mutants Generation} and that to automate the transformation from original source code to a mutant. Table \ref{tab:Path mutators} lists all the path-sensitive mutational operators which we used to develop mutators using TXL transformation rules. Each version of these path-sensitivity mutators will inject one change in each SmartThings base source code app.

When mutants are generated for path-sensitivity, a mutant meta-data repository is also created to track how many mutants were caught and killed by the taint flow analysis tools used for the evaluation. This meta-data repository will keep the information about the path-sensitive mutants. Listings \ref{list:preference}-\ref{list:Mutated code path} demonstrates the mutational injection process on a SmartThings app. The mutator used in this example demonstrates the \textit{Aps1\_1} (Add path sensitivity) mutational operator.

To begin, the mutator will search in the preference block of the SmartThings App to find and extract one sensitive source from the input section:\\\\\\

\begin{lstlisting}[style=Groovy, caption={Preference}, label=list:preference]
preferences {
  section("When all of these people leave home") {
    input ~"presence"~, "capability.presenceSensor", multiple: true 
  }
}
\end{lstlisting}

So, from Listing \ref{list:preference} it extracts the input variable \textit{\$presence}. Afterwards, it finds a random function to inject a code block with an if-else path containing sensitive source \textit{\$presence} and a sink. For \textit{Aps1\_1} the sink used is \textit{sendSms}. The created block for \textit{Aps1\_1} will also add a line to define a variable \textit{phone} which stores a phone number to be used as the first parameter in the \textit{sendSms} sink function. An example injection vector is presented in Listing \ref{list:Mutated code path}:
\begin{lstlisting}[style=Groovy, caption = {Mutated code}, label = list:Mutated code path]
~def phone = "11111111111"~
def message = *"benign info"*
if(x)
  message = `"${presence}"`
else
  message = *"benign info"*
~sendSMS("1111111111", messages)~
\end{lstlisting}

This code, when properly analysed, will produce two possible outputs, one vulnerable, and that if the condition in the if statement evaluates to true, and one benign output which is generated if the condition evaluates to false. A precise static analysis tool should report the two paths recognizing which one is vulnerable and which one is benign.

In the following subsections, we describe each path-sensitivity mutator, their different versions and demonstrate their use on taint flow analysis tools.
\subsubsection{Add Path-sensitivity (Aps)}
This mutator is designed to add a designated block to create path sensitivity in the application code. We have designed six different variants of this mutational operator. Three of them are designed with sink \textit{sendSms} and three with \textit{httpPost}. Since \textit{sendPush} is very similar to \textit{sendSms} we did not use it for this operator. 
\begin{figure*}[!ht]
	\centerline{\includegraphics[width=.99\textwidth]{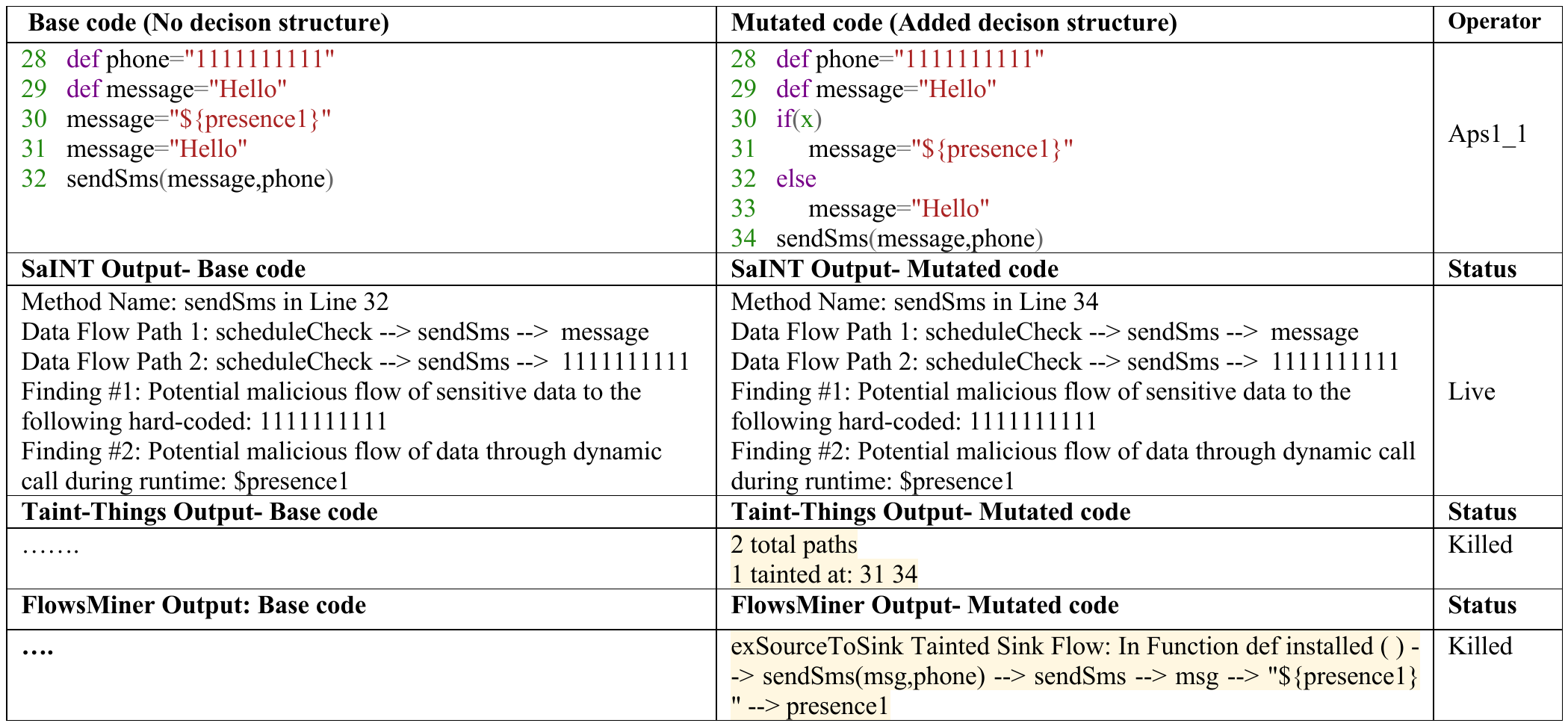}}
	\caption{Add path sensitivity with vulnerable data in the if clause and an SMS sink (Aps1\_1) Operator}
	\label{fig:Aps1_1_updated}
\end{figure*}

For all of the variants of the operators, we created the base version of this code by adding a code block without decision structure (i.e if-else statement). In every base version, the base code statements will be in the exact same order as the mutated one except the added if-else statement. For each of the sinks, we did one version where the first path, the if clause, had the vulnerable information that can be leaked through a sensitive sink which executes after the if-else block. Listing \ref{list:Mutated code path} demonstrate this case.
\begin{figure*}[!ht]
	\centerline{\includegraphics[width=.99\textwidth]{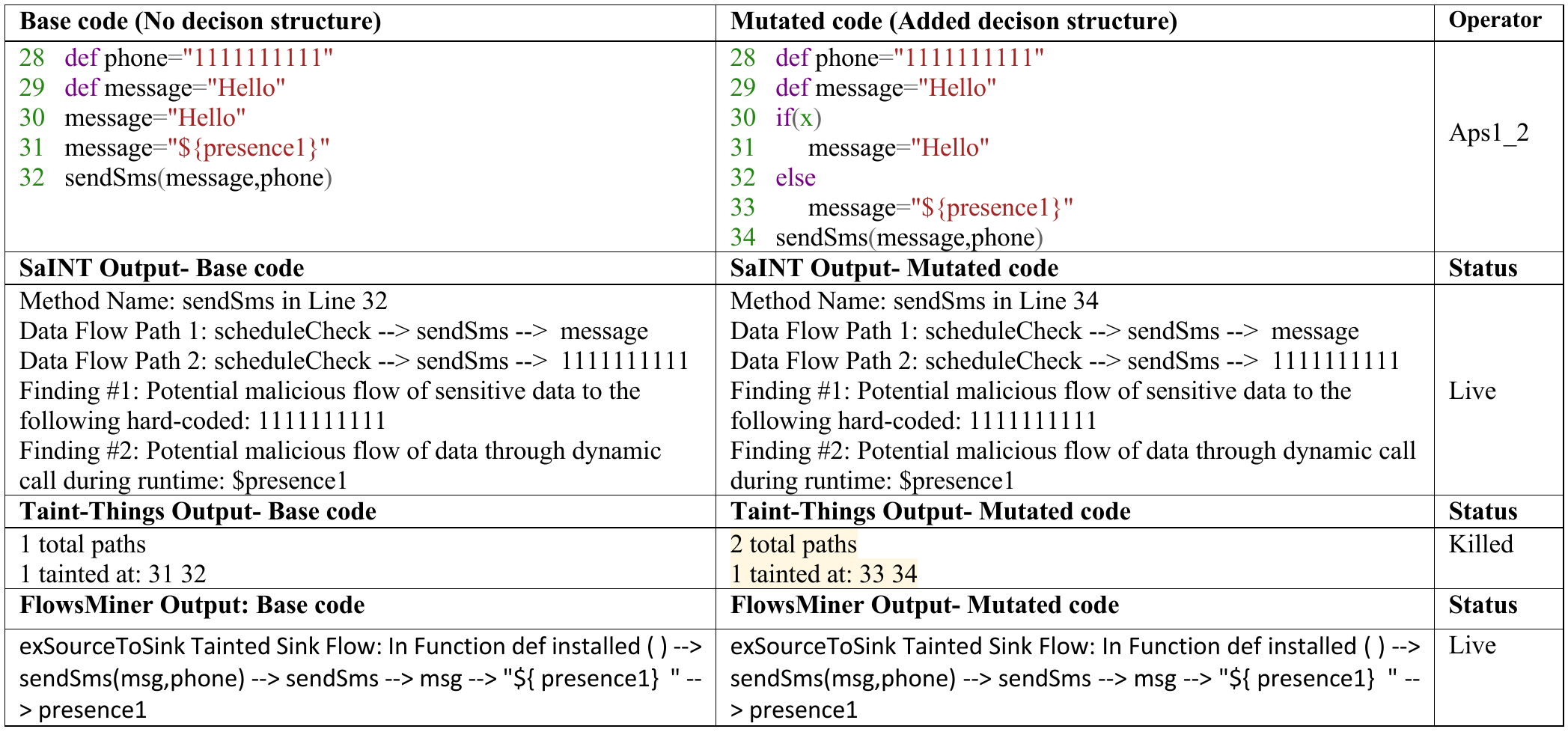}}
	\caption{Add path sensitivity with vulnerable data in the else clause and an SMS sink  (Aps1\_2) Operator}
	\label{fig:Aps1_2_updated}
\end{figure*}
\begin{figure*}[!ht]
	\centerline{\includegraphics[width=.99\textwidth]{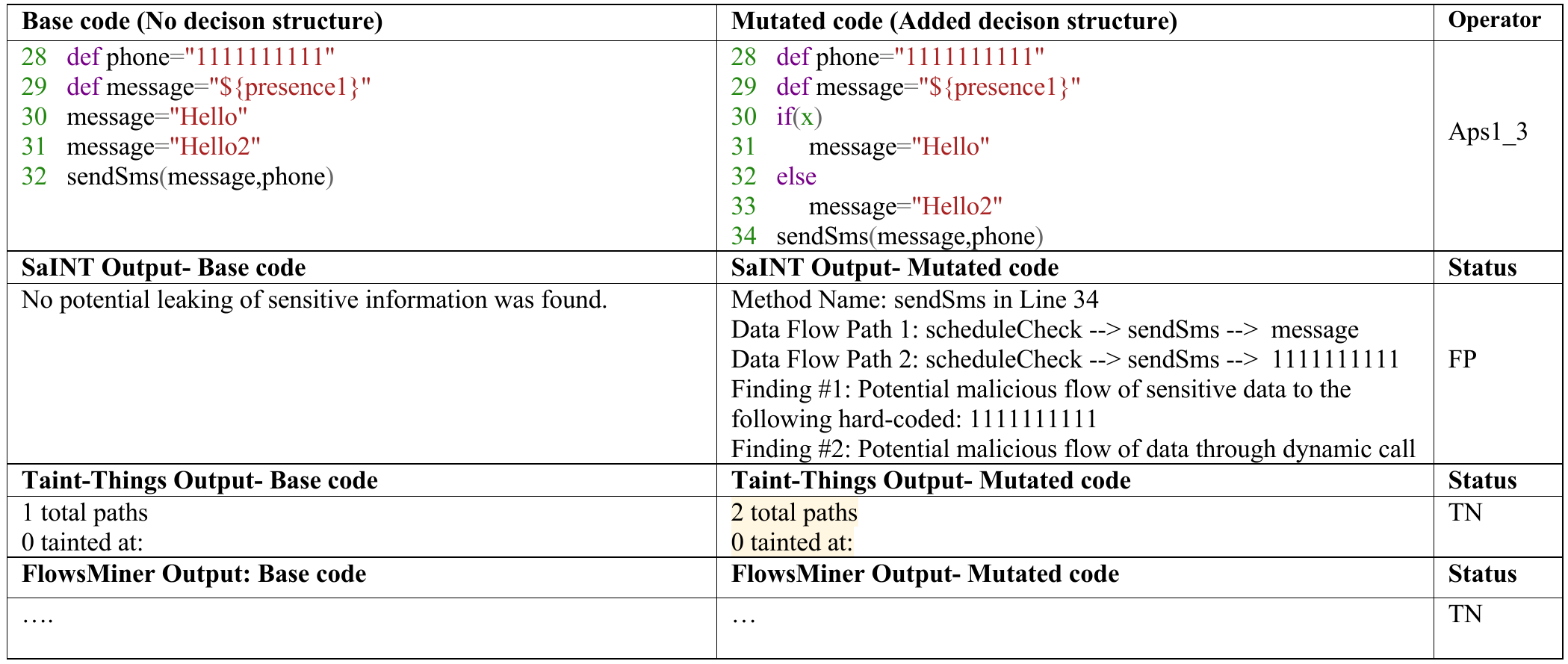}}
	\caption{Add path sensitivity with benign data and an SMS sink (Aps1\_3) Operator}
	\label{fig:Aps1_3_updated}
\end{figure*}

In the if clause path in \textit{Aps1\_1}, as demonstrated in Figure \ref{fig:Aps1_1_updated}, the value of the message is replaced by the vulnerable value of input \textit{\$presence1}. Before the if clause, it had a benign value. Also, the else clause contains a benign value as well. In the base version of the mutator, there will be no if-else statements, so the first value of the message is benign. Then it is replaced by \textit{\$presence1} and again, it is replaced by a benign value. Ultimately making it a benign one. The second version will have vulnerable information in the else clause path. This version is demonstrated in \textit{Aps1\_2}, shown in Figure \ref{fig:Aps1_2_updated}. 

The third and last version will be a benign one. Before the if-else block, it will have a vulnerable source, but its value will be replaced in both if and else clauses. So ultimately whichever path is executed the vulnerable information cannot be leaked. \textit{Aps1\_3}, shown in Figure \ref{fig:Aps1_3_updated}, and \textit{Aps2\_3}, not shown, are the benign version for SMS and HTTP sink. Before the if-else block, the message and takeParams contain the value that is associated with an inputted value, making them vulnerable. But both if and else clauses replace it with something benign that has no relation with the inputted value. They are created as an equivalent mutants; equivalent in terms of security vulnerability, since both are benign and both are used to evaluate the false positives reported by the security analysis tools.

 For \textit{Aps1\_1},\textit{Aps1\_2}, \textit{Aps2\_1} and \textit{Aps2\_2}, SaINT fails to distinguish between the base code and the mutated code. In \textit{Aps1\_1} and \textit{Aps2\_1} the base code is benign and the mutated code is malicious for one path. In  \textit{Aps1\_2} and \textit{Aps2\_2}, the base code is malicious and the mutated code is path sensitively malicious. For all these four versions, SaINT and FlowsMiner identify all of the codes as the same level of malicious and the tool's finding does not differ from base code to mutated code, keeping the mutation alive. On the other hand, Taint-Things can identify the changes from base code to mutated code whether it is changing from benign to malicious or malicious to path sensitively malicious. Thus, it is able to kill them all. For \textit{Aps1\_3} and \textit{Aps2\_3}, both the base code and mutated code are benign. Here, Taint-Things and FlowsMiner report that the base code has one path and that the mutated code has two paths with zero tainted flows, making the result a true negative where SaINT marks both of apps as malicious resulting in \hlc[highlight]{a} false positive.

\subsubsection{Add HTTP-Request Path-sensitivity (AHps)}
This operator works on an existing path in the code. Instead of using a \emph{sendSms}, The operator uses \emph{httpPost} sink and the data is sent over the internet. From the original code, presented in Listing \ref{list:Mutated AHPs code path_updated}, we have designed four mutators, three mutators generates vulnerable mutant and one generates a benign mutant. Listing \ref{list:Mutated AHPs1 code path_updated} demonstrates the \textit{AHps1} operator to generate a vulnerable mutant. The mutator adds a vulnerable sensitive source, \textit{\$motion1}, as part of the message constructed to be sent/leaked via the internet. A precise path sensitive static analysis tool will generate 2 paths from this code, reporting one as vulnerable, and that if the \textit{if condition} evaluates to true. However, If the condition evaluates to false, the executed else path will generate benign output. The other three operators \textit{AHps1\_1}, \textit{AHps2}, \textit{AHps3} generates the other 2 vulnerable and one benign mutants as follows:
\begin{itemize}
\item \textit{AHps1\_1}: The operator is quite similar to \textit{AHps1} operator, presented above. The only difference is, this version doesn't require an existing else clause. Here, the mutator declares a benign variable before an if block. It then changes it's value to a vulnerable inside the if block. When the if condition evaluates to false, the variable remains benign. The mutator send the variable through the Httpr Post sink after the if statement executes.  

\item \textit{AHps2}: Unlike \textit{AHps1\_1}, This operator alters the vulnerability in the else block. 
\item \textit{AHps3}: This operator generates a benign mutant. Before the if-else block, one variable is created with a vulnerable value. Inside both the if and else clauses, it is replaced by a harmless value. At the end, when the variable is passed to the sink, regardless of which path is executed, a non-malicious value will be sent. This is an equivalent mutant, as both the base and mutated code is benign but it will help us evaluate the false positives.
\end{itemize}
\begin{lstlisting}[style=Groovy, float, caption={Original code to demonstrate three Add HTTP-Request Path-sensitivity (AHps) versions}, label=list:Mutated AHPs code path_updated]
if (elapsed >= threshold) {
  log.debug "turning off lights"
  switches.on()
  state.inactiveAt = null
} else {
  log.debug "${elapsed / 1000} sec since motion stopped"
}
\end{lstlisting}
For the first three category of mutants, the mutants will be killed or kept alive, but for this case, it will either be reported as a false positive or a true negative. When evaluated on the tools, SaINT and FlowsMiner have failed to identify the differences between the flows. They produces the same result for both base and mutated code, keeping the mutant alive. SaINT also fails to identify \textit{AHps3} mutant correctly, making a false-positive result, while FlowsMiner identifies it correctly as benign. On the other hand, Taint-Things can distinguish the paths, resulting in killing the first three mutants (\textit{AHps1},\textit{AHps1\_1}, \textit{AHps2}) and producing a true negative result for \textit{AHps3} .

\begin{lstlisting}[style=Groovy, float, caption={Add HTTP-Request Path-sensitivity with vulnerability in the if clause (AHps1)}, label=list:Mutated AHPs1 code path_updated]
~def takeParams = [ uri:"https://attacker.com", path:"",~
      ~requestContentType:"application/x-www-form-urlencoded",~
      ~body:[ *"massage":"hello"* ] ]~
      
if (elapsed >= threshold) {
  ~takeParams = [uri:"https://attacker.com",path:"",~
  ~requestContentType:"application/x-www-form-urlencoded",~
  ~body:[`"massage":"${motion1}"` ]]~
  log.debug "turning off lights"
  switches.on()
  state.inactiveAt = null 
} else {
  ~takeParams = [uri:"https://attacker.com", path:"",~
  ~requestContentType:"application/x-www-form-urlencoded",~
  ~body:[*"massage":"Hello"*]]~
  log.debug "${elapsed / 1000} sec since motion stopped" 
}

~try {~
  httpPost(takeParams) {resp->
  ~if(resp.status == 200){~
    ~log.debug"attack succeeded"~
  ~} else{~
    ~ log.error"attack failed"}~
  ~}~ 
~} catch(Exception e) {~
 ~log.error"Unexpected exception",e~
~}~

\end{lstlisting}

\begin{lstlisting}[style=Groovy, float, caption={Original code to demonstrate Add Push-message Path-sensitivity (APps)}, label=list:orgial APps code path_updated]
if (!d) 
  httpError(404, "Device not found")
else 
  device(d, name)
\end{lstlisting}

\subsubsection{Add Push-message/send-message Path-sensitivity (APps/AMps):}
Theses two operators are constructed in the same way as AHps. However, the sensitive sink is different. For AMps the sink is  \emph{sendSms}, and \emph{sendPush} for APps. From an original code such as Listing \ref{list:orgial APps code path_updated}, each of \emph{APps} and \emph{AMps} mutational operators generate four different variants: three with vulnerability and one benign. 
\begin{lstlisting}[style=Groovy, float, caption={Mutated code to demonstrate Add Push-message Path-sensitivity (APps)}, label=list:Mutated APps code path_updated]
*def msg ="hello"*
if (!d) {
   *msg = "hello"*
   httpError(404, "Device not found")
} else {
  `msg = "$switches"`
  device(d, name)
}
`sendPush(msg)`
\end{lstlisting}    
When evaluated on the tools, they behave exactly the same as those mutants generated from AHps operators and demonstrated in the previous subsection. An example vulnerable mutant generated by the APps mutator is demonstrated in Listing \ref{list:Mutated APps code path_updated}. 

\subsection{Path Sensitive Mutants' Experiment and Results}\label{sec:Path Sensitive mutants' experiment and result}
For the path sensitivity, we generated the maximum amount of \hlc[highlight] {mutants}. We have four operators for path sensitivity, where each operator has four to six versions and each version has two files each. So, from our dataset of 90 benign \hlc[highlight] {apps}, 87 \hlc[highlight] {apps} were able to produce mutants. For these 87 \hlc[highlight] {apps}, each of them generates 6, 9 or 18 mutants. Most of them produce 18 mutants. The mutation happens based on the code. If it does not match a certain pattern, it will not generate a version for it. That results in producing 1890 mutants. We can check  the generated file numbers of each type of operator of path sensitivity from Table \ref{tab:Files created for each mutational operators}. This is a huge number to run for SaINT where this tool has an over quota and server connection issues from time to time. Also, Taint-Things takes more time running path sensitive analysis. So to minimize our problem we ended up randomly selecting 30 out of the 87 benign \hlc[highlight] {apps} which were capable of generating mutants. To make the dataset diverse, we took one \hlc[highlight] {app from} the SaINT benchmark apps, four \hlc[highlight] {apps from} the official marketplace apps, four \hlc[highlight] {apps from} the third party apps, and 21 \hlc[highlight] {apps from} on the official community apps. The distribution of the chosen dataset is shown in Table \ref{tab:files and mutation executed in each tool for each version of path sensitivity}. 

\begin{table}[!ht]
\centering
\caption{\label{tab:files and mutation executed in each tool for each version of path sensitivity} Mutated files executed for different versions of mutators in path-sensitivity}
\begin{tabular}{|l|l|l|l|l|l|}
\hline
Mutator               & \begin{tabular}[c]{@{}l@{}}Mutator\\ version/\\ dataset\end{tabular} & \begin{tabular}[c]{@{}l@{}}SaINT\\ (files)\end{tabular} & \begin{tabular}[c]{@{}l@{}}Forum\\ (files)\end{tabular} & \begin{tabular}[c]{@{}l@{}}Market\\ (files)\end{tabular} & \begin{tabular}[c]{@{}l@{}}Community\\ (files)\end{tabular} \\ \hline
                      & \begin{tabular}[c]{@{}l@{}}No of\\ \hlc[highlight] {Apps} in\\ Dataset\end{tabular}   & 1                                                       & 4                                                       & 4                                                        & 21                                                          \\ \hline
\multirow{6}{*}{Aps}  & Aps1\_1                                                              & 2                                                       & 8                                                       & 8                                                        & 42                                                          \\ \cline{2-6} 
                      & Aps1\_2                                                              & 2                                                       & 8                                                       & 8                                                        & 42                                                          \\ \cline{2-6} 
                      & Aps1\_3                                                              & 2                                                       & 8                                                       & 8                                                        & 42                                                          \\ \cline{2-6} 
                      & Aps2\_1                                                              & 2                                                       & 8                                                       & 8                                                        & 42                                                          \\ \cline{2-6} 
                      & Aps2\_2                                                              & 2                                                       & 8                                                       & 8                                                        & 42                                                          \\ \cline{2-6} 
                      & Aps2\_3                                                              & 2                                                       & 8                                                       & 8                                                        & 42                                                          \\ \hline
\multirow{4}{*}{AMps} & AMps1                                                                & 2                                                       & 8                                                       & 8                                                        & 32                                                          \\ \cline{2-6} 
                      & AMps1\_1                                                             & 2                                                       & 8                                                       & 8                                                        & 38                                                          \\ \cline{2-6} 
                      & AMps2                                                                & 2                                                       & 8                                                       & 8                                                        & 32                                                          \\ \cline{2-6} 
                      & AMps3                                                                & 2                                                       & 8                                                       & 8                                                        & 32                                                          \\ \hline
\multirow{4}{*}{APps} & APps1                                                                & 2                                                       & 8                                                       & 8                                                        & 32                                                          \\ \cline{2-6} 
                      & APps1\_1                                                             & 2                                                       & 8                                                       & 8                                                        & 38                                                          \\ \cline{2-6} 
                      & APps2                                                                & 2                                                       & 8                                                       & 8                                                        & 32                                                          \\ \cline{2-6} 
                      & APps3                                                                & 2                                                       & 8                                                       & 8                                                        & 32                                                          \\ \hline
\multirow{4}{*}{AHps} & AHps1                                                                & 2                                                       & 8                                                       & 8                                                        & 32                                                          \\ \cline{2-6} 
                      & AHps1\_1                                                             & 2                                                       & 8                                                       & 8                                                        & 38                                                          \\ \cline{2-6} 
                      & AHps2                                                                & 2                                                       & 8                                                       & 8                                                        & 32                                                          \\ \cline{2-6} 
                      & AHps3                                                                & 2                                                       & 8                                                       & 8                                                        & 32                                                          \\ \hline
\end{tabular}
\end{table}

\subsubsection{Evaluating Path Sensitivity}
To evaluate path sensitivity, we consider the change in the results from the base to the mutated code. If a tool provides a different result, showing consideration for the execution paths, we consider the tool to provide a path sensitive analysis. For the three tools tested, only Taint-Things provides an option for a detailed result for all the possible paths.

For SaINT, all of the mutants were alive and the benign equivalent mutants were identified as malicious, resulting into a false positive report. In Table \ref{tab:Path sensitivity result}, in the SaINT result column, we indicated the false positive as \textit{FP} and live as \textit{L} after the number. While FlowsMiner keeps the mutants alive, it accurately detects \textit{Aps1\_3}, \textit{Aps2\_3}, \textit{AMps3}, \textit{APps3} and \textit{AHps3 } as true negatives.

\begin{table}
\centering
\caption{\label{tab:Path sensitivity result}Path sensitivity result for mutator and different versions}
\begin{tabular}{|l|l|l|l|l|l|}
\hline
Mutator               & \begin{tabular}[c]{@{}l@{}}Mutator\\ version/\end{tabular} & Mutant & \begin{tabular}[c]{@{}l@{}}SaINT\\ (Result)\end{tabular} & \begin{tabular}[c]{@{}l@{}}Taint-Things\\ (Result)\end{tabular} & \begin{tabular}[c]{@{}l@{}}FlowsMiner\\ (Result)\end{tabular}  \\ \hline
\multirow{6}{*}{Aps}  & Aps1\_1                                                    & 30     & 30L                                                    & 30K    & 30L                                                        \\ \cline{2-6} 
                      & Aps1\_2                                                    & 30     & 30L                                                    & 30K     & 30L                                                       \\ \cline{2-6} 
                      & Aps1\_3                                                    & 30     & 30FP                                                   & 30TN      & 30TN                                                       \\ \cline{2-6} 
                      & Aps2\_1                                                    & 30     & 30L                                                    & 30K     & 30L                                                        \\ \cline{2-6} 
                      & Aps2\_2                                                    & 30     & 30L                                                    & 30K      & 30L                                                       \\ \cline{2-6} 
                      & Aps2\_3                                                    & 30     & 30FP                                                   & 30TN       & 30TN                                                      \\ \hline
\multirow{4}{*}{AMps} & AMps1                                                      & 25     & 25L                                                    & 25K          & 25L                                                 \\ \cline{2-6} 
                      & AMps1\_1                                                   & 28     & 28L                                                    & 28K    & 28L                                                        \\ \cline{2-6} 
                      & AMps2                                                      & 25     & 25L                                                    & 25K    & 25L                                                         \\ \cline{2-6} 
                      & AMps3                                                      & 25     & 25FP                                                   & 25TN     & 25TN                                                      \\ \hline
\multirow{4}{*}{APps} & APps1                                                      & 25     & 25L                                                    & 25K      & 25L                                                        \\ \cline{2-6} 
                      & APps1\_1                                                   & 28     & 28L       & 28L                                               & 28K                                                           \\ \cline{2-6} 
                      & APps2                                                      & 25     & 25L                                                    & 25K       & 25L                                                       \\ \cline{2-6} 
                      & APps3                                                      & 25     & 25FP                                                   & 25TN    & 25TN                                                         \\ \hline
\multirow{4}{*}{AHps} & AHps1                                                      & 25     & 25L                                                    & 25K          & 25L                                                  \\ \cline{2-6} 
                      & AHps1\_1                                                   & 28     & 28L                                                    & 28K     & 28L                                                        \\ \cline{2-6} 
                      & AHps2                                                      & 25     & 25L                                                    & 25K     & 25L                                                       \\ \cline{2-6} 
                      & AHps3                                                      & 25     & 25FP                                                   & 25TN    & 25TN                                                        \\ \hline
\end{tabular}
\end{table}

For Taint-things, all the mutants were killed for all three types and the benign equivalent mutants were identified as true negative. The true negative is denoted as \textit{TN}  after the digit in the Taint-Things results column. The \textit{K} after the digit means killed and the number before that indicates how many were killed.

For this case, if we calculate the recall and precision for all of these mutators, the result is shown in Table \ref{tab:Precision and recall for 2 files-path}. Though there are false positive identified for SaINT, it does not affect the precision and recall as it is zero.
For Taint-Things the precision and recall both are a hundred percent. We can conclude from these results that SaINT does not opt for path sensitive analysis, rather it does a path insensitive analysis to ensure security. And though the path sensitive version of Taint-Things shows the great result for recall and precision, it was slower than before in its execution. But in this evaluation experiment, we did not include any performance analysis based on time.

\begin{table}[!ht]
\centering
\caption{\label{tab:Precision and recall for 2 files-path}Precision and recall for path mutators (2 files counted as one mutant) for SaINT, Taint-Things and FlowsMiner}
\begin{tabular}{llllllll}
\hline
{Type}             & {Mutators} & \multicolumn{2}{l}{SaINT} & \multicolumn{2}{l}{Taint-Things} & \multicolumn{2}{l}{FlowsMiner} \\ \cline{3-8} 
                                  &                           & Rec.\%        & Prec.\%       & Rec.\%            & Prec.\%        & Rec.\%            & Prec.\%     \\ \hline
{Path-Sensitivity}           
                                  & Aps                      & 00         & 00         & 100             & 100    & 00         & 00            \\
                                  & AMps                      & 00         & 00         & 100             & 100    & 00         & 00            \\
                                  & APps                      & 00         & 00         & 100             & 100    & 00         & 00            \\
                                  & AHps                      & 00         & 00         & 100             & 100      & 00         & 00    
                                  
                                  \\ \cline{2-8} 
                                  
                                                                    & Overall                      & 00         & 00         & 100             & 100      & 00         & 00    
                                                                    \\ \cline{2-8}

\end{tabular}

\vspace{-0.1 cm}
\end{table}

\subsubsection{Evaluating the Effect of Path Sensitivity:}
In the previous subsection, we focused on evaluating whether the tools provides any sort of path sensitive analyses, and thus the criteria was the change in the result from the base code to the mutator. But if we want to consider the effect of path sensitivity analysis on the accurate detection of sensitive data leaks we also need to look into the correctness of the results. To calculate this and its effect on precision and recall, we consider the total possible paths. So, for a mutator with a single if-statement or an if-else-statement, we get 2 paths which are tested. For example, Listing \ref{list:IfElse} shows a program with two paths, one that is tainted when the if condition is false and one which is benign if  the condition is true. For this evaluation, we consider that testing this program should give us the two results that should be recognized, one benign path and one that is tainted.

\begin{lstlisting}[style=Groovy, float=tb, caption={If-else Statement with Two Paths}, label=list:IfElse]
def message = "This contains $sensitiveData"
if (x) {
  message = "This is benign"
}
sendSms(message)
\end{lstlisting}

In the generated path mutants, we split the mutants into three groups. \textit{Group 1}, containing if-else-statements with 1 benign path and 1 tainted. This group includes Aps1\_1, Aps1\_2, Aps2\_1, Aps2\_2, AMps2, APps1, Apps2, AMps1, AHps1, and Ahps2 . The second group, \textit{Group 2}, containing if-statements with 1 benign path and 1 tainted, which includes AMps1\_1, APps1\_1, and AHps1\_1. For both groups SAINT and FlowsMiner only reports the tainted flow. We consider that they're not reporting the benign path as a false positive and thus giving us 100\% recall 50\% precision. While Taint-Things identifies the two paths and reports one as tainted, the other as benign with 100\% recall and precision.

Finally, \textit{Group 3} containing if-else-statements that have 2 benign paths. This includes Aps1\_3, Aps2\_3, AMps3, APps3 and AHps3. SaINT reports a false positive result, and thus provides 0\% precision. FlowsMiner reports the program as benign, so we consider this as a true negative and gives us 100\% precision. Taint-Things identifies the two paths and reports them as benign with 100\% precision. The results and their distribution for each mutator is show in Table \ref{tab:Precision and recall Path Effect_}.

\begin{table}[!ht]
	\centering
	\caption{\label{tab:Precision and recall Path Effect_} Precision and recall for path mutators if we consider the possible paths in the files }
	\begin{tabular}{llllllll}
		\hline
		{Type}             & {Mutators} & \multicolumn{2}{l}{SaINT} & \multicolumn{2}{l}{Taint-Things} & \multicolumn{2}{l}{FlowsMiner} \\ \cline{3-8} 
		  &      & Rec.\% & Prec.\% & Rec.\% & Prec.\% & Rec.\% & Prec.\% \\ \hline
		{Path-Sensitivity}           
		  & Aps  & 100     & 33.3      & 100    & 100     & 100     & 66.6      \\
		  & AMps & 100     & 38      & 100    & 100     & 100     & 62      \\
		  & APps & 100     & 38      & 100    & 100     & 100     & 62      \\
		  & AHps & 100     & 38      & 100    & 100     & 100     & 62      
		                                  
		\\ \cline{2-8} 
          & Overall & 100     & 37      & 100    & 100     & 100     & 63      
        \\ \cline{2-8} 
	        
	\end{tabular}
	
	\vspace{-0.1 cm}
\end{table}


\section{Context-sensitive Mutation Generation and Analysis}\label{ch: context -Automated Mutants Generation}

Our framework presents three new mutational operators for context-sensitivity. Same as flow and path sensitive mutational operators, we utilized three common sinks of IoT SmartThings to outline the operators. 
Table \ref{tab:context mutators} defines the proposed context-sensitive mutational operators. 
\subsection{Context-sensitive Mutants Generation}\label{sec:How the context-sensitive Mutants are generated}
The process of automated mutant generation for context sensitivity is similar to flow and path sensitivity. As mentioned in Section \ref{ch: Flow -Automated Mutants Generation}  and Section \ref{Path Automated Mutants Generation}, we use TXL source transformation technology \cite{cordy2009excerpts} and a modified version of TXL Groovy grammar developed by Schmeidl et al. \cite{schemild}. The transformation process is dependent on the TXL Groovy grammar, it takes SmartThings apps as input, which are written in Groovy, and then parse and transform those apps into mutants using a set of transformation rules designed to cover a set of mutational operators. Table \ref{tab:context mutators} presents the set of proposed context-sensitive mutational operators. 
Each mutator, will apply one mutational operator, to inject one probable fault/change in each SmartThings \hlc[highlight] {app}. In the case of context-sensitivity mutaional operators, the goal is to generate mutants that can expose scenarios to evaluate whether static analysis provides context sensitivity analysis or not.  

\begin{table*}[!ht]
\centering
\caption{\label{tab:context mutators} Proposed context-sensitive mutational operators}
\begin{tabular}{|l|l|l|l|} 
\hline
\textbf{ Name}  & \textbf{~ ~ ~ ~ ~ ~ ~ ~ ~ ~ ~ ~ ~ ~ ~ ~ ~ ~ ~Description }                                                                                                                                 & \textbf{Type }  \\ 
\hline
\textit{AMcs }  & \begin{tabular}[c]{@{}l@{}}Add a send message sink to leak information if we found\\ one function is called from multiple context(the sink will \\be added to be reached from only one context) \end{tabular} & Context         \\ 
\hline
\textit{APcs }  & \begin{tabular}[c]{@{}l@{}}Add Push sink to leak information if we found one function\\ is called from multiple context(the sink will be added to be\\ reached from only one context) \end{tabular}           & Context         \\ 
\hline
\textit{AHcs }  & \begin{tabular}[c]{@{}l@{}}Add HTTP sink to leak information if we found one function\\ is called~from multiple context(the sink will be added to be\\ reached from only one context) \end{tabular}           & Context         \\
\hline

\end{tabular}
\end{table*}
A mutant meta-data repository is created in this stage to keep track of information regarding the context-sensitive mutations which help us later for statistical analysis. Listings \ref{list:preference1}-\ref{list:Mutated code} demonstrate the context-sensitive mutational injection process on a SmartThings app. The mutator used in this example demonstrates the injection process for the Add Message context sensitivity (\textit{AMcs}) mutational operator.
To begin, the mutator will pattern match the preference block in the SmartThings app to find and extract one sensitive source as defined in the input section shown in Listing \ref{list:preference1} as an example.

\begin{lstlisting}[style=Groovy, float, caption={Preference}, label=list:preference1]
preferences {
  section("When all of these people leave home") {
    input ~"people"~, "capability.presenceSensor", multiple: true 
  }
}
\end{lstlisting}

\vspace{0.7 cm}

Afterwards, it will pattern match an assignment statement with a function call where the value returned from that function call is assigned to a variable. The pattern of the function call we are looking for is highlighted in Listing \ref{list:Function Call}. 
 
\begin{lstlisting}[style=Groovy, caption={Original- Benign code}, label=list:Function Call]
def  validateCommand(device, deviceType, command) {
  /*any other lines  */  
  ~def capabilityCommands = getDeviceCapabilityCommands(deviceType)~
  /*any other lines  */
}

/*any other functions  */ 

def getDeviceCapabilityCommands(deviceCapabilities) { 
  def map = [:]
  deviceCapabilities.collect {
    map[it.name] = it.commands.collect{it.name.toString()}
  }
  return map 
}
\end{lstlisting}

The \textit{AMcs} mutator will then construct the injection vector. First, it clones the function \textit{getDeviceCapabilityCommands (type)}. When cloning, it adds an additional argument to the cloned function where the value of that variable is also added to the \textit{return} statement in the function's body. Second, The mutator,  replaces the function body of the original function \textit{getDeviceCapabilityCommands (type)} with benign statements as shown in Listing \ref{list:Mutated code}. 

In the \textit{installed()} function, the mutator adds one call to the original function and one call to the added cloned function and assign their values to two different variables.  When adding the function call, the mutator is designed to only pass a benign value as parameter to the original function. In the cloned function, in addition to the  benign parameter passed to the original function, the mutator adds a new parameter which uses the sensitive data extracted from the Input section of the SmartThings app under-analysis, as demonstrated in Listing \ref{list:preference1}. 
 We make sure only one parameter is vulnerable and that to create the simplest context-sensitive scenario. 

After adding the calls in \textit{installed} function, the mutator adds two statements to construct a sensitive sink that will leak sensitive data. It adds a variable containing a phone number and then add a sink \textit{sendSMS()}, which takes as parameter the variable \textit{functionCall2}, which is supposed to receive the malicious returned value from the cloned function and send it to a phone\# stored in the first parameter \textit{phone}. Listing \ref{list:Mutated code} reflects the changes. 

\begin{figure*}[!ht]
	\centerline{\includegraphics[width=.99\textwidth]{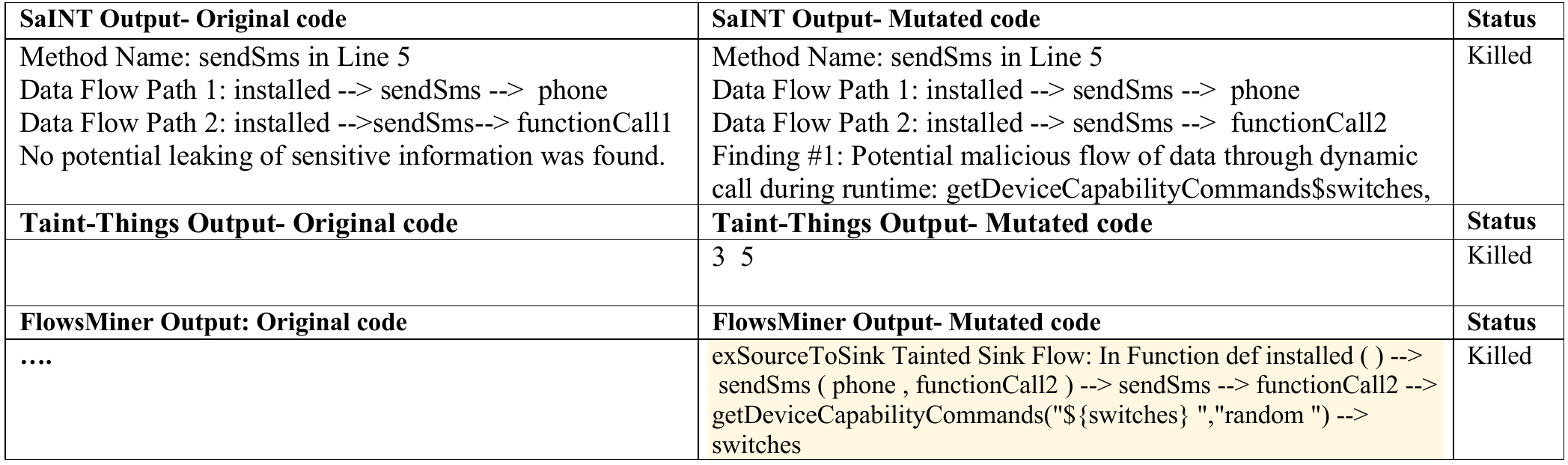}}
	\caption{Add Message Context-sensitivity (AMcs) Operator}
	\label{fig:AMcs}
\end{figure*}

If we want to create a benign version for \textit{AMcs}, then instead of passing \textit{functionCall2}, \textit{Listing \ref{list:Mutated code} line 5},we pass the benign variable \textit{functionCall1}. This variable stores a benign value returned from a benign function. The parameter it passes is also benign. Figure \ref{fig:AMcs} demonstrates that SaINT, Taint-Things and FlowsMiner were able to kill mutants generated from the \textit{AMcs} operator.

In this subsection we described the first context sensitivity operator, AMcs, in the following subsections we describe each context-sensitive mutational operator and the corresponding mutants generated from the original code shown in Listing \ref{list:Function Call}.

\begin{lstlisting}[style=Groovy, float, caption={Mutated Code}, label=list:Mutated code]
def installed() {
  ~def functionCall1 = getDeviceCapabilityCommands("random")~
  *def functionCall2 = getDeviceCapabilityCommands(`"${switches}"`,"random")*
  ~def phone = "11111111111"~
  ~sendSms(phone, functionCall2)~
  /*any other lines  */ 
}

/*any other functions  */ 

def getDeviceCapabilityCommands(deviceCapabilities) {
  `return true`
}

*def getDeviceCapabilityCommands(`flag`,deviceCapabilities) {*
  `return flag`
  def map = [:]
  deviceCapabilities.collect{
    map[it.name] = it.commands.collect{it.name.toString()}
  }
  return map
}
\end{lstlisting}

\subsubsection{Add Push-message Context-sensitivity (APcs)}
APcs follows the same pattern as AMcs. The only difference is that it adds a \textit{sendPush} sink instead of a \textit{sendSms} sink and adds the function calls in the \textit{updated} function instead of \textit{installed}. Because it uses the sendPush sink, it does not need a variable to store a phone number. The \textit{updated} function is another built-in function that is called by SmartThings framework, so additional context will also be avoided when using this function. The results of SaINT, Taint-Things and FlowsMiner are also the same. They were able to kill the mutation for the example file. 
    
\subsubsection{Add HTTP Context-sensitivity (AHcs)}
\begin{figure*}[!ht]
	\centerline{\includegraphics[width=.99\textwidth]{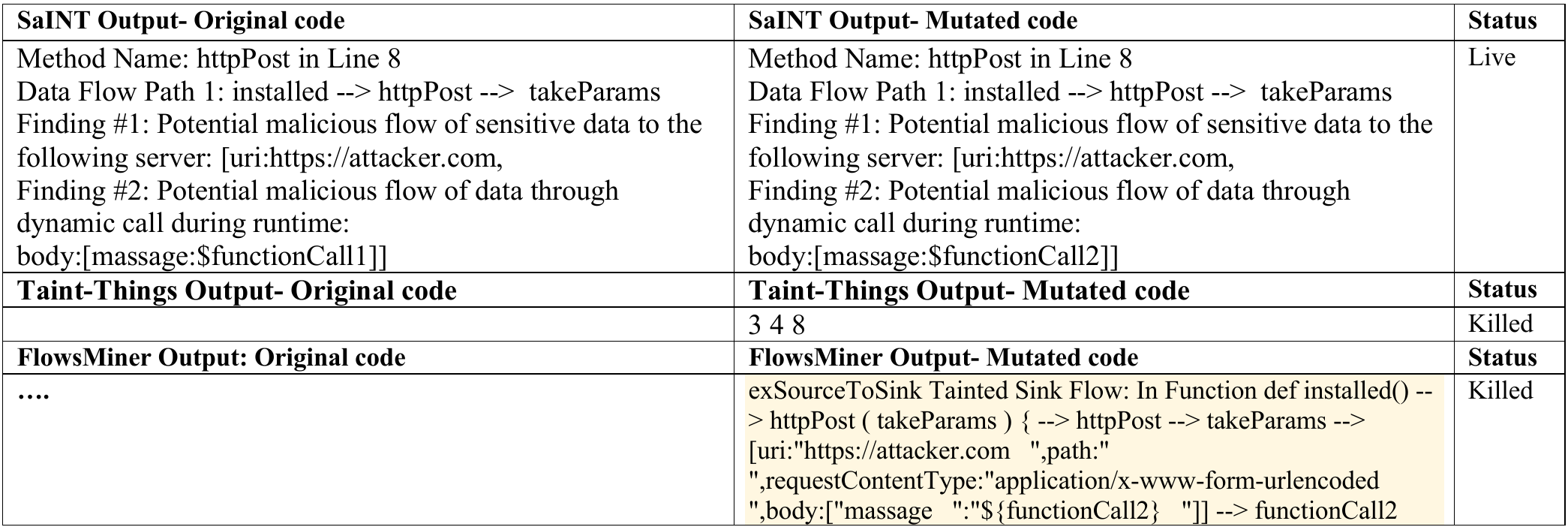}}
	\caption{Add HTTP Context-sensitivity (AHcs) Operator}
	\label{fig:AHcs}
\end{figure*}

AHcs follows the same pattern as AMcs but uses the \textit{httpPost} sink. Because of this, some of the added code  differ, since \textit{httpPost} is slightly different. We add the sink \textit{httpPost} with a parameter \textit{takeParams}. In the base version, \textit{takeParams} does not take any vulnerable variable, but in the mutated version, \textit{takeParams} takes a vulnerable input related variable. \textit{httpPost} also does not need a variable to store phone numbers. For the result, SaINT's did not preform the same as AMcs; it was able identify which variable is passed to the sink, but it marks both of the paths as vulnerable. On the other hand, Taint-Thing and FlowsMiner were able to kill the mutant in this example. Figure \ref{fig:AHcs} demonstrates an example of such operator.

\begin{lstlisting}[style=Groovy, float, caption={Mutated AHcs Code}, label=list:Mutated code2]
def installed() {
  def functionCall1 = getDeviceCapabilityCommands("random")
  ~def functionCall2 = getDeviceCapabilityCommands(`"${switches}"`,"random")~
  def takeParams = [  uri:"https://attacker.com",
  path:"",
  ~requestContentType: "application/x-www-form-urlencoded",~
  ~body:["massage": `"${functionCall2}"`] ]~ 
  ~try {~  
    ~httpPost(takeParams){ resp->~
      ~if (resp.status==200) {~
        ~ log.debug"attack succeeded"~
      } else { 
        ~log.error"attack failed"~
      }
    }    
  ~} catch(Exception e) {~
    ~log.error"Unexpected exception",e ~ 
  }    
}

% omitted statements

def getDeviceCapabilityCommands(deviceCapabilities) {
  *return true* 
}

~def getDeviceCapabilityCommands(flag,deviceCapabilities) {~
  `return flag`
  def map = [:]
  deviceCapabilities.collect {
    map[it.name] = it.commands.collect {it.name.toString()}
  }
  return map
}
\end{lstlisting}  

\subsection{Context-Sensitive Mutants Experiment and Result}\label{sec:Context Sensitive mutants' experiment and result}
In this section, we focus on the category of context-sensitivity mutational operators. We developed three contexts- sensitivity mutational operators for this framework. For each original benign \hlc[highlight] {app}, we generate one base file and one mutant, both containing the changes described previously in this section. Together, they are considered as one mutation. We generated 3 mutants, with 6 files, for the marketplace apps in the data-set and 13 mutants, with 26 files, for community apps. We show the results in Table \ref{tab:Mutants Killed context sensitive}, and that for both cases: considering the two files as one mutant and each file as one. 

In Table \ref{tab:Mutants Killed context sensitive}, the letter after the digit shows the status. \textit{K} stands for Killed, \textit{L} stands for Live, and \textit{FP} stands for False Positive, \textit{TN} stands for True Negative.  Then, we  discuss the results when considering the two files as one mutant. For AMcs and APcs, SaINT was able to perfectly classify the files and distinguish between the context, which resulted in all of the mutants being killed, but for AHcs, even though it was able to distinguish which variable is passed, it failed to classify the files correctly. It classified both contexts as malicious. Thus, all the AHcs mutants were alive in this experiment for SaINT. Most of the mutants generated are classified and distinguished correctly by Taint-Things and FlowsMiner.


\begin{table}
\centering
\caption{\label{tab:Mutants Killed context sensitive}Mutants result for context sensitivity}
\resizebox{\linewidth}{!}{%
\begin{tabular}{|l|l|l|l|l|l|l|l|l|l|} 
\hline
\begin{tabular}[c]{@{}l@{}}Mutator\\ (2files \\ creating\\ one \\ mutation) \end{tabular} & Mutant              & \begin{tabular}[c]{@{}l@{}}SaINT\\ Result \end{tabular} & \begin{tabular}[c]{@{}l@{}}Taint-\\ Things\\ Result \end{tabular} & \begin{tabular}[c]{@{}l@{}}FlowsMiner\\Result\end{tabular} & \begin{tabular}[c]{@{}l@{}}Mutator \\ version\\ (Each file one \\ mutant) \end{tabular} & Mutant & \begin{tabular}[c]{@{}l@{}}SaINT\\ Result \end{tabular} & \begin{tabular}[c]{@{}l@{}}Taint-\\ Things\\ Result \end{tabular} & \begin{tabular}[c]{@{}l@{}}FlowsMiner\\Result\end{tabular}  \\ 
\hline
\multirow{2}{*}{AMcs}                                                                     & \multirow{2}{*}{16} & \multirow{2}{*}{16K}                                    & \multirow{2}{*}{16K}                                              & \multirow{2}{*}{16K}                                       & AMcscontext1                                                                            & 16     & 16TN                                                    & 15TN,1FP                                                          & 16TN                                                        \\ 
\cline{6-10}
                                                                                          &                     &                                                         &                                                                   &                                                            & AMcscontext2                                                                            & 16     & 16K                                                     & 16K                                                               & 16K                                                         \\ 
\hline
\multirow{2}{*}{APcs}                                                                     & \multirow{2}{*}{16} & \multirow{2}{*}{16K}                                    & \multirow{2}{*}{16K}                                              & \multirow{2}{*}{16K}                                       & APcscontext1                                                                            & 16     & 16TN                                                    & 15TN,1FP                                                          & 16TN                                                        \\ 
\cline{6-10}
                                                                                          &                     &                                                         &                                                                   &                                                            & APcscontext2                                                                            & 16     & 16K                                                     & 16K                                                               & 16K                                                         \\ 
\hline
\multirow{2}{*}{AHcs}                                                                     & \multirow{2}{*}{16} & \multirow{2}{*}{16L}                                    & \multirow{2}{*}{16K}                                              & \multirow{2}{*}{16K}                                       & AHcscontext1                                                                            & 16     & 16FP                                                    & 15TN,1FP                                                          & 16TN                                                        \\ 
\cline{6-10}
                                                                                          &                     &                                                         &                                                                   &                                                            & AHcscontext2                                                                            & 16     & 16K                                                     & 16K                                                               & 16K                                                         \\
\hline
\end{tabular}
}
\caption*{K = Killed, L = Lived, FP = False Positive, TN = True Negative}
\end{table}

Taint-Things was able to identify all the injected changes from the base apps and the mutants. But it had more reported vulnerabilities in the benign and mutated files generated from one original app. All the three benign mutants that Taint-Things failed to identify as benign share the same original file which it marks as tainted due to its aggressive handling of state variables. The app made extensive use of them which made Taint-Things mark multiple methods within as tainted. Listing \ref{list:tainted-file}  shows a function with a sink call made using a state variable. This would be tagged as tainted by Taint-Things due to how it generally handles all state variables as potentials sources, while SAINT and FlowsMiner identify no of sensitive information leaking from that call

\begin{lstlisting}[style=Groovy, float=ht, caption={File Marked as Taitned by Taint-Things}, label=list:tainted-file]
def locationEventHandler(evt) {
  state.locationSubscriptionMap[location.id].each {
  //...
    httpPostJson(params) {
      //.. 
    }	
  //...
  }
}
\end{lstlisting}
So the precision and recall rate of SaINT for AMcs and APcs is 100\% but for AHcs it is 0\%. The precision and recall rate of Taint-Things for each operator when the benign file is counted as base is 100\% for both. This is shown in Table \ref{tab:Precision and recall for 2 files-Context} as it can differentiate the change . If we consider each file as a mutant, then the result is shown in Table \ref{tab:Precision and recall for 1 file-Context}. Here, the precision rate of Taint-Things dropped to 96.88 for classifying one file wrong for each type. While FlowsMiner managed to get 100\% Precision and Recall for both cases.
  
For Context sensitivity, we see a drop in the precision rate for Taint-Things, even though it is still over 95 percent. For SaINT, we see that it can differentiate the context but still can have trouble identifying which programs are malicious or benign based on the sink. While FlowsMiner didn't have a problem handling the mutators.
 

\begin{table}[!ht]
	\centering
	\caption{\label{tab:Precision and recall for 2 files-Context}Precision and recall for context mutators (Two files counted as one mutant) for SaINT and Taint-Things}
	
	\begin{tabular}{llllllll}
		\hline
		{Type}             & {Mutators} & \multicolumn{2}{l}{SaINT} & \multicolumn{2}{l}{Taint-Things} & \multicolumn{2}{l}{FlowsMiner}\\ \cline{3-8} 
		  &      & Rec.\% & Prec.\% & Rec.\% & Prec.\% & Rec.\% & Prec.\% \\ \hline
		{Context-Sensitivity}           
		  & AMcs & 100    & 100     & 100    & 100     & 100    & 100     \\
		  & APcs & 100    & 100     & 100    & 100     & 100    & 100     \\
		  & AHcs & 00     & 00      & 100    & 100     & 100    & 100     
		                                  
		\\ \cline{2-8} 
		  & Overall & 66.6 & 66.6 & 100    & 100     & 100    & 100     
	\end{tabular}
	
	\vspace{-0.1 cm}
\end{table}
 
\begin{table}[!ht]
\centering
\caption{\label{tab:Precision and recall for 1 file-Context}Precision and recall for context mutators (one file counted as one mutant) for SaINT and Taint-Things}

\begin{tabular}{llllllll}
\hline
{Type}             & {Mutators} & \multicolumn{2}{l}{SaINT} & \multicolumn{2}{l}{Taint-Things} & \multicolumn{2}{l}{FlowsMiner} \\ \cline{3-8} 
                                  &                           & Rec.\%        & Prec.\%       & Rec.\%            & Prec.\%      & Rec.\%            & Prec.\%      \\ \hline
{Context-Sensitivity}           
                                  & AMcs                      & 100         & 100         & 96.88             & 100  & 100  & 100            \\
                                  & APcs                      & 100         & 100         & 96.88             & 100            & 100  & 100  \\
                                  & AHcs                      & 100         & 50         & 96.88             & 100  & 100  & 100      
                                  
                                  \\ \cline{2-8} 
                 & Overall                      & 100         & 83.3         & 96.88             & 100  & 100  & 100                           
                                                                    \\ \cline{2-8} 

\end{tabular}

\label{Approach}
\vspace{-0.1 cm}
\end{table}  


\section{Discussion}\label{sec: discussion}

In this paper, \hlc[highlight]{we try to answer two research questions: 
\textit{RQ1.} How can we quantitatively evaluate static taint analysis tools in IoT to ensure security?  and \textit{RQ2.} Is the proposed methodology able to differentiate the security analysis tools and identify their weakness?

To answer RQ1 we developed a a mutational framework that is capable of generating test cases for static analysis tools and} we demonstrate how the proposed framework is useful in assessing  the available tools, marking their strengths and shortcomings \hlc[highlight]{to answer RQ2}. While the original available datasets used for assessing the analyses sensitivities did not show much change, using the mutational testing provided more clear assessment of the tools' performance. It also tests the  consistency of the tool's results over a large number of test cases.

Using our framework, we have quantitatively evaluated three tools,( SaINT, Taint-Things and FlowsMiner), on their sensitivity analysis for tainted flow identification.

For flow-sensitive mutators, we calculated the precision and recall in two ways: 

First, we generate a benign and a vulnerable flow for each \hlc[highlight] {app} in the dataset, and we consider each of them as a separate mutant where the benign ones would be classified as equivalent mutants and would be used to check for the false-positive rates in the tools' reports. We found that  SaINT, Taint-Things and FlowsMiner have a recall rate of 100\% for this calculation. But SaINT's precision rate dropped to 50\% where Taint-Things' precision remains 100\%. 
    
Second, we chose to consider the generated benign mutant as an original or a base code which we used to generate the vulnerable mutant and then we checked whether the two tools can differentiate between the flows (benign vs venerable) mutants. This will directly confirm whether the tested tool uses a flow-sensitive analysis or not. With recall and precision rate of zero, SaINT gets confirmed for using a flow insensitive analysis technique, while scoring 100\% for both, Taint-Things and FlowsMiner are confirmed to use flow-sensitive analysis. 
    
For path-sensitivity mutators, we designed four versions; one that takes a benign path as a base and changes it to a mutant containing a path-sensitive malicious one; two versions that take a malicious base and makes the vulnerability path sensitive; and finally a version that produces an equivalent benign mutant from a benign base. We specifically designed them in such a way to check that the tools are using path sensitive analysis. For SaINT and FlowMiners, we got zero percent recall and precision. SaINT also had false positives, on the other hand, FlowsMiners avoided giving that false positive case. The result confirms that SaINT and FlowsMiner are  not using path sensitive analysis. Taint-Things got recall and precision rate of 100\% here, which proves that it uses path sensitive analysis technique. 

We also provided a test to evaluate the impact of path sensitivity by checking the correctness of the tool's reports. We dealt with each path as an expected result to be reported and tested which tools report them. We've seen that SaINT and FlowMiner were able to identify the tainted results, giving a recall of 100\% but had their precision drop by ignoring the potential benign paths if another tainted path existed. And while SaINT also reported a false positive in mutants that had two benign paths, FlowMiner accurately reported the app as a true negative. Taint-Things on the other hand is able to give a detailed report of all the potential paths

When it comes to context-sensitivity mutators, the results confirmed that the tools are using context-sensitive analysis up to a certain level. For SaINT, even after getting the context, it failed to differentiate benign from malicious when using a specific sink.

If we give equal weight to each of the three categories of mutators that address the three levels of sensitivity analysis, we can average their results to get an estimated overall precision and recall for each of the tools. When one file is counted as one mutant and the correctness of the path sensitivity results are considered SaINT has 100\% recall and 56.8\% precision, Taint-Things has 99\% recall and 100\% precision, and FlowsMiner has 100\% recall and 87.6\% precision. This is shown in Table \ref{tab:overall precision and recall}.  If we consider 2 files as one, showing whether the tool uses sensitive analysis or not, SaINT has 25\% recall and 25\% precision, Taint-Things has 100\% recall and 100\% precision, and FlowsMiner has 66.6\% recall and 66.6\% precision.

\begin{table}[!h]
	\centering
	\caption{\label{tab:overall precision and recall} Overall precision and recall if we count one file as one mutant}
	\begin{tabular}{lllllll}
		\hline
		{Type}             & \multicolumn{2}{l}{SaINT} & \multicolumn{2}{l}{Taint-Things} & \multicolumn{2}{l}{FlowsMiner} \\ \cline{2-7} 
		   & Rec.\% & Prec.\% & Rec.\% & Prec.\% & Rec.\% & Prec.\% \\ \hline
		{Flow-Sensitivity}           
		  & 100     & 50     & 100    & 100     & 100     & 100      \\
		{Path-Sensitivity}       
		   & 100     & 37      & 100    & 100     & 100     & 63      \\
		{Context-Sensitivity}       
		  & 100     & 83.3      & 96.88    & 100     & 100     & 100     
		\\ \cline{1-7} 
		{Overall}     
          & 100     & 56.8      & 99    & 100     & 100     & 87.6      
        \\ \cline{1-7} 
	        
	\end{tabular}
	
	\vspace{-0.1 cm}
\end{table}

For Taint-Things, it can distinguish the change from the created base file to the generated mutant. But it failed to identify the mutants generated from one app that contained an extensive usage of state variable which marked it aggressively as a potential source. It failed for all the benign equivalent mutants generated from one source \hlc[highlight] {app} when we only had sixteen source \hlc[highlight] {apps}. On the other hand, FlowsMiner handled the provided files without an issue. To understand the exception SaINT had for context-sensitivity when a specific sink is used, we need more operators with different sinks, whereas to understand Taint-Things' exception when identifying benign mutants generated from a specific app, we need to generate more mutants from different apps. For context-sensitivity analyses, we can say that the tools implement it but SaINT and Taint-Things have certain limitations. 

Each of the tested tools used a different approach for static analysis. SaINT makes an intermediate representation of the code and analyzes it as a call graph. Taint-Things performs the analysis directly on the code using code transformation and provides additional sensitive analysis as modules. FlowsMiner uses text mining to detect the tainted flows. And while the tools' results varied on our tests, it might be hard to determine whether that is due to the technique used or the specific implementation of the tool. A bigger survey with a larger group of tools might be needed to examine a correlation between different techniques and their performance in the tests. \hlc[highlight]{It should also be noted that we test for two things, the correctness in detecting the tainted flows and the ability to provide sensitive analyses by detecting the difference in the output. Cases where the tools times out or do not run on certain mutants are excluded in our calculation of precision and recall, since they do not give us relevant data. Nonetheless, this is relevant if one is comparing the general performance of the tools.}

\section{Future Work}
We have proposed 13 unique mutational operators for three different kinds of sensitivity analysis, however, new operators can be added easily to the framework to address different kinds of sensitivity analysis. This is an on-going effort and our goal is to extend the framework to include all possible mutational operators and to conduct a large scale experiment to qualitatively evaluate state-of-the-art tools in other domains such as Android. \hlc[highlight]{In the future, we envision the following work:}
    \begin{itemize}
        \item First, we plan to extend the framework to evaluate the security analysis tools of SmartThings for other types of security vulnerabilities.
        \item Second, we can expand this framework to other platforms of IoT as they share some similarities.
        
        \item Third, we can explore how the basic idea of the framework can be used or extended to analyze security analysis tools in different domains and platforms other than IoT.

        \item Fourth, the proposed framework can be extended to address a different kind of sensitivity analysis such as field sensitivity. For example, IoT devices use state and atomicState objects. A field-sensitive analysis is required to track all fields defined in the state and atomicState objects \cite{celik2018program}. A field insensitive analysis will fail to distinguish between the fields. New operators can be added to addressing leakage related field sensitivity.
    
        \item Fifth, while the framework  only deals with evaluating tools for tainted-flows sensitivity analysis in a single file, this can be extended to multi-file applications for other platforms. The extended framework  will be able to insert one fault throughout the application and will be able to test the security analysis tools better where needed.
        
        \item Finally, since SmartThings apps are single-file apps, in general, and the evaluated tools analyze single-file apps, the scope of research of this paper is on mutation testing for single-file apps. Expanding the framework to handle multi-file apps might be more relevant in other platforms which may have more multi-file apps, such as Android applications. However, such extensions to the framework will need to explore other categories of vulnerabilities that are caused by interacting apps, or interaction between multiple files within the application. This would be an immediate future work that would require another set of tools to evaluate.

 \end{itemize}

\section{ Conclusions}\label{sec: Conclusions}
    
We have presented an automated evaluation framework that artificially creates and injects known security vulnerabilities to the application under test. The framework can be used to accurately measure the recall and precision of taint flow analysis tools for IoT apps. We proposed 13 unique mutational operators to evaluate security analysis tools in IoT platforms which can be extended to any other platform as well. These mutational operators, or mutators, are designed based on three different kinds of sensitive analyses. With the help of these operators, we can evaluate if a tool  is able to provide different kinds of sensitive analyses and up to which level it is capable of doing that. We also used the classical problem of equivalent mutants to our advantage by measuring false positive using them. 

We proposed and implemented a framework that can evaluate taint flow analysis tools and with the conducted experiment, we evaluated three tools quantitatively. By adding better quality operators, this tool can become stronger in evaluation. One novel idea that is produced by our work is using the equivalent mutants to evaluate false positives. Equivalent mutants are always treated as a problem, while we were able to use it to our advantage. This idea can be implemented in any other evaluation tool as well.

\ack This work is supported in part by the Natural Sciences and Engineering Research Council of Canada (NSERC), Grant No.    RGPIN/06283-2018

\bibliographystyle{wileyj}
\bibliography{References}   
\clearpage

\end{document}